\documentclass[conference,a4paper]{IEEEtran}

\usepackage[utf8]{inputenc} 
\usepackage[T1]{fontenc}
\usepackage{url}              
\usepackage{cite}             
\usepackage[cmex10]{amsmath}  
\usepackage{mleftright}       
\mleftright                   

\usepackage{graphicx}         
\usepackage{booktabs}         

\usepackage{latexsym}
\usepackage{amssymb}
\usepackage{comment}
\usepackage{cite}
\usepackage{url}

\global\long\def\L{\mathsf{L}}
\global\long\def\D{\mathsf{D}}
\global\long\def\Sgen{\mathcal{S}_\mathsf{gen}}
\global\long\def\Kgen{\mathcal{K}_\mathsf{gen}}
\global\long\def\A{\mathcal{A}}

\newcommand{\calVarX}{{\cal X}}

\newcommand{\calVarZ}{{\cal Z}}

\newcommand{\ovK}{\overline{K}}
\newcommand{\ovZ}{\overline{Z}}
\newcommand{\ovKZ}{\overline{K}\,\overline{Z}}

\begin{document}

\title{
Universal Coding for Shannon Ciphers 
under Side-Channel Attacks
}
\author{%
	\IEEEauthorblockN{Yasutada Oohama and Bagus Santoso}
	\IEEEauthorblockA{University of Electro-Communications, Tokyo, Japan\\ 
	Email: \url{{oohama,santoso.bagus}@uec.ac.jp}}
%
}
\maketitle

\newcommand{\Zaas}{
\title{
Universal Source Encryption 
\\$\quad$ under Side-Channel Attacks} 

\author{%
  \IEEEauthorblockN{Yasutada Oohama}
  \IEEEauthorblockA{%
    Please do NOT provide authors' names and affiliations\\
    in the paper submitted for review, but keep this placeholder.\\
    ISIT23 follows a \textbf{double-blind reviewing policy}.}
}

\maketitle
}
\begin{abstract}
We study the universal coding problem for the general framework
of source encryption with a symmetric key under the side-channel
attacks, which is posed and investigated by Oohama and Santoso (2022). 
The reliable and secure rate region indicating 
the trade off between the compression rate of the ciphertext
and the rate constraint imposed on the adversary for secure 
source transmission was established by Oohama and Santoso (2022). 
In this paper we focus on our attention to strengthening 
the direct coding theorem. We prove the existence of encryption/decryption schemes, 
which are universal in the sense that they work effectively  
for {\it any distributions of the plain text}, {\it any noisy 
channels through which the adversary observe the corrupted version
of the key}, and any measurement device used for collecting 
the physical information. Those schemes have a good performance
such that if we compress the ciphertext 
with rate within the reliable and secure rate region, 
then: (1) anyone with secret key will be able to decrypt and decode the ciphertext correctly, but (2) any adversary who obtains the ciphertext 
and also the side physical information will not be able to obtain 
any information about the hidden source as long as the 
leaked physical information is encoded with a rate within the rate constraint.
\end{abstract}

\newcommand{\BagusIsitabst}{
\begin{abstract}
	We are interested in investigating the security of source encryption
	with a symmetric key under side-channel attacks.
	In this paper, we propose a general framework of source encryption
	with a symmetric key under the side-channel attacks,
	which applies to \emph{any} source encryption
	with a symmetric key and \emph{any} kind of side-channel attacks
    targeting the secret key. We also propose a new security criterion for
	strong secrecy under side-channel attacks,
	which is a natural extension of mutual information, i.e.,
	\emph{the maximum conditional mutual information between the plaintext
	and the ciphertext given the adversarial key leakage,
	where the maximum is taken over all possible plaintext distribution}.
	Under this new criterion, we successfully formulate the rate region,
	which serves as both necessary and sufficient conditions to have
	secure transmission even under side-channel attacks.
	Furthermore, we also prove another theoretical result on 
	our new security criterion, which might be interesting in its own right:
    in the case of the discrete memoryless source, 
    no perfect secrecy under side-channel attacks in the
    standard security criterion, i.e., the ordinary mutual information,
    is achievable  without achieving perfect
	secrecy in this new security criterion,
	although our new security criterion is more strict than
	the standard security criterion.
\end{abstract}
}

\newcommand{\qed}{\hfill$\square$}
\newcommand{\suchthat}{\mbox{~s.t.~}}
\newcommand{\markov}{\leftrightarrow}

\newcommand{\argmax}{\mathop{\rm argmax}\limits}
\newcommand{\argmin}{\mathop{\rm argmin}\limits}

\newcommand{\ExP}{\rm e}

\newcommand{\Cls}{class NL}
\newcommand{\vSpa}{\vspace{0.3mm}}
\newcommand{\Prmt}{\zeta}
\newcommand{\pj}{\omega_n}

\newfont{\bg}{cmr10 scaled \magstep4}
\newcommand{\bigzerol}{\smash{\hbox{\bg 0}}}
\newcommand{\bigzerou}{\smash{\lower1.7ex\hbox{\bg 0}}}
\newcommand{\nbn}{\frac{1}{n}}
\newcommand{\ra}{\rightarrow}
\newcommand{\la}{\leftarrow}
\newcommand{\ldo}{\ldots}
\newcommand{\typi}{A_{\epsilon }^{n}}
\newcommand{\bx}{\hspace*{\fill}$\Box$}
\newcommand{\pa}{\vert}
\newcommand{\ignore}[1]{}


\newtheorem{proposition}{Proposition}
\newtheorem{definition}{Definition}
\newtheorem{theorem}{Theorem}
\newtheorem{lemma}{Lemma}
\newtheorem{corollary}{Corollary}
\newtheorem{remark}{Remark}
\newtheorem{property}{Property}

\newcommand{\defeq}{:=}

\newcommand{\Qed}{\hbox{\rule[-2pt]{3pt}{6pt}}}
\newcommand{\beq}{\begin{equation}}
\newcommand{\eeq}{\end{equation}}
\newcommand{\beqa}{\begin{eqnarray}}
\newcommand{\eeqa}{\end{eqnarray}}
\newcommand{\beqno}{\begin{eqnarray*}}
\newcommand{\eeqno}{\end{eqnarray*}}
\newcommand{\ba}{\begin{array}}
\newcommand{\ea}{\end{array}}

\newcommand{\vc}[1]{\mbox{\boldmath $#1$}}
\newcommand{\lvc}[1]{\mbox{\scriptsize \boldmath $#1$}}
\newcommand{\svc}[1]{\mbox{\scriptsize\boldmath $#1$}}

\newcommand{\wh}{\widehat}
\newcommand{\wt}{\widetilde}
\newcommand{\ts}{\textstyle}
\newcommand{\ds}{\displaystyle}
\newcommand{\scs}{\scriptstyle}
\newcommand{\vep}{\varepsilon}
\newcommand{\rhp}{\rightharpoonup}
\newcommand{\cl}{\circ\!\!\!\!\!-}
\newcommand{\bcs}{\dot{\,}.\dot{\,}}
\newcommand{\eqv}{\Leftrightarrow}
\newcommand{\leqv}{\Longleftrightarrow}

\newcommand{\irr}[1]{{\color[named]{Red}#1\normalcolor}}		
\newcommand{\irWh}[1]{{\color[named]{White}#1\normalcolor}}

\newcommand{\hugel}{{\arraycolsep 0mm
                    \left\{\ba{l}{\,}\\{\,}\ea\right.\!\!}}
\newcommand{\Hugel}{{\arraycolsep 0mm
                    \left\{\ba{l}{\,}\\{\,}\\{\,}\ea\right.\!\!}}
\newcommand{\HUgel}{{\arraycolsep 0mm
                    \left\{\ba{l}{\,}\\{\,}\\{\,}\vspace{-1mm}
                    \\{\,}\ea\right.\!\!}}
\newcommand{\huger}{{\arraycolsep 0mm
                    \left.\ba{l}{\,}\\{\,}\ea\!\!\right\}}}
\newcommand{\Huger}{{\arraycolsep 0mm
                    \left.\ba{l}{\,}\\{\,}\\{\,}\ea\!\!\right\}}}
\newcommand{\HUger}{{\arraycolsep 0mm
                    \left.\ba{l}{\,}\\{\,}\\{\,}\vspace{-1mm}
                    \\{\,}\ea\!\!\right\}}}

\newcommand{\hugebl}{{\arraycolsep 0mm
                    \left[\ba{l}{\,}\\{\,}\ea\right.\!\!}}
\newcommand{\Hugebl}{{\arraycolsep 0mm
                    \left[\ba{l}{\,}\\{\,}\\{\,}\ea\right.\!\!}}
\newcommand{\HUgebl}{{\arraycolsep 0mm
                    \left[\ba{l}{\,}\\{\,}\\{\,}\vspace{-1mm}
                    \\{\,}\ea\right.\!\!}}
\newcommand{\hugebr}{{\arraycolsep 0mm
                    \left.\ba{l}{\,}\\{\,}\ea\!\!\right]}}
\newcommand{\Hugebr}{{\arraycolsep 0mm
                    \left.\ba{l}{\,}\\{\,}\\{\,}\ea\!\!\right]}}
\newcommand{\HUgebr}{{\arraycolsep 0mm
                    \left.\ba{l}{\,}\\{\,}\\{\,}\vspace{-1mm}
                    \\{\,}\ea\!\!\right]}}

\newcommand{\hugecl}{{\arraycolsep 0mm
                    \left(\ba{l}{\,}\\{\,}\ea\right.\!\!}}
\newcommand{\Hugecl}{{\arraycolsep 0mm
                    \left(\ba{l}{\,}\\{\,}\\{\,}\ea\right.\!\!}}
\newcommand{\hugecr}{{\arraycolsep 0mm
                    \left.\ba{l}{\,}\\{\,}\ea\!\!\right)}}
\newcommand{\Hugecr}{{\arraycolsep 0mm
                    \left.\ba{l}{\,}\\{\,}\\{\,}\ea\!\!\right)}}

\newcommand{\hugepl}{{\arraycolsep 0mm
                    \left|\ba{l}{\,}\\{\,}\ea\right.\!\!}}
\newcommand{\Hugepl}{{\arraycolsep 0mm
                    \left|\ba{l}{\,}\\{\,}\\{\,}\ea\right.\!\!}}
\newcommand{\hugepr}{{\arraycolsep 0mm
                    \left.\ba{l}{\,}\\{\,}\ea\!\!\right|}}
\newcommand{\Hugepr}{{\arraycolsep 0mm
                    \left.\ba{l}{\,}\\{\,}\\{\,}\ea\!\!\right|}}

\newcommand{\MEq}[1]{\stackrel{
{\rm (#1)}}{=}}

\newcommand{\MLeq}[1]{\stackrel{
{\rm (#1)}}{\leq}}

\newcommand{\ML}[1]{\stackrel{
{\rm (#1)}}{<}}

\newcommand{\MGeq}[1]{\stackrel{
{\rm (#1)}}{\geq}}

\newcommand{\MG}[1]{\stackrel{
{\rm (#1)}}{>}}

\newcommand{\MPreq}[1]{\stackrel{
{\rm (#1)}}{\preceq}}

\newcommand{\MSueq}[1]{\stackrel{
{\rm (#1)}}{\succeq}}

\newcommand{\MSubeq}[1]{\stackrel{
{\rm (#1)}}{\subseteq}}

\newcommand{\MSupeq}[1]{\stackrel{
{\rm (#1)}}{\supseteq}}

\newcommand{\MRarrow}[1]{\stackrel{
{\rm (#1)}}{\Rightarrow}}

\newcommand{\MLarrow}[1]{\stackrel{
{\rm (#1)}}{\Leftarrow}}

\newcommand{\SZZpp}{
}

\newcommand{\vcc}{{c}^n}
\newcommand{\vck}{{k}^n}
\newcommand{\vcx}{{x}^n}
\newcommand{\vcy}{{y}^n}
\newcommand{\vcz}{{z}^n}
\newcommand{\vckone}{{k}_1^n}
\newcommand{\vcktwo}{{k}_2^n}
\newcommand{\vcxone}{{x}^n}

\newcommand{\vcxtwo}{{x}_2^n}
\newcommand{\vcyone}{{y}_1^n}
\newcommand{\vcytwo}{{y}_2^n}

\newcommand{\cvcx}{\check{x}^n}
\newcommand{\cvcy}{\check{y}^n}
\newcommand{\cvcz}{\check{z}^n}
\newcommand{\cvcxone}{\check{x}^n}
\newcommand{\cvcxtwo}{\check{x}_2^n}

\newcommand{\hvcx}{\widehat{x}^n}
\newcommand{\hvcy}{\widehat{y}^n}
\newcommand{\hvcz}{\widehat{z}^n}
\newcommand{\hvckone}{\widehat{k}_1^n}
\newcommand{\hvcktwo}{\widehat{k}_2^n}

\newcommand{\hvcxone}{\widehat{x}^n}
\newcommand{\hvcxtwo}{\widehat{x}_2^n}

\newcommand{\lvcc}{{c}^n}
\newcommand{\lvck}{{k}^n}
\newcommand{\lvcx}{{x}^n}
\newcommand{\lvcy}{{y}^n}
\newcommand{\lvcz}{{z}^n}

\newcommand{\lvckone}{{k}_1^n}
\newcommand{\lvcktwo}{{k}_2^n}
\newcommand{\lvcxone}{{x}_1^n}
\newcommand{\lvcxtwo}{{x}_2^n}
\newcommand{\lvcyone}{{y}_1^n}
\newcommand{\lvcytwo}{{y}_2^n}

\newcommand{\hlvckone}{\widehat{k}_1^n}
\newcommand{\hlvcktwo}{\widehat{k}_2^n}
\newcommand{\hlvcxone}{\widehat{x}^n}
\newcommand{\hlvcxtwo}{\widehat{x}_2^n}

\newcommand{\lcvcxone}{\check{x}^n}
\newcommand{\lcvcxtwo}{\check{x}_2^n}

\newcommand{\rvcc}{{C}^n}
\newcommand{\rvck}{{K}^n}
\newcommand{\rvcx}{{X}^n}
\newcommand{\rvcy}{{Y}^n}
\newcommand{\rvcz}{{Z}^n}
\newcommand{\rvccone}{{C}_1^n}
\newcommand{\rvcctwo}{{C}_2^n}
\newcommand{\rvckone}{{K}_1^n}
\newcommand{\rvcktwo}{{K}_2^n}
\newcommand{\rvcxone}{{X}^n}

\newcommand{\rvcxtwo}{{X}_2^n}
\newcommand{\rvcyone}{{Y}_1^n}
\newcommand{\rvcytwo}{{Y}_2^n}
\newcommand{\hrvcx}{\widehat{X}^n}
\newcommand{\hrvcxone}{\widehat{X}_1^n}
\newcommand{\hrvcxtwo}{\widehat{X}_2^n}
\newcommand{\crvcx}{\check{X}^n}
\newcommand{\crvcxone}{\check{X}_1^n}
\newcommand{\crvcxtwo}{\check{X}_2^n}

  \newcommand{\cvcc}{\check{c}^m}
 \newcommand{\lcvcc}{\check{c}^m}
 \newcommand{\crvcc}{\check{C}^m}
\newcommand{\lcrvcc}{\check{C}^m}
\newcommand{\lrvcc}{{C}^n}
\newcommand{\lrvck}{{K}^n}
\newcommand{\lrvcx}{{X}^n}
\newcommand{\lrvcy}{{Y}^n}
\newcommand{\lrvcz}{{Z}^n}
\newcommand{\lrvckone}{{K}_1^n}
\newcommand{\lrvcktwo}{{K}_2^n}
\newcommand{\lrvcxone}{{X}_1^n}
\newcommand{\lrvcxtwo}{{X}_2^n}
\newcommand{\lrvcyone}{{Y}_1^n}
\newcommand{\lrvcytwo}{{Y}_2^n}
\newcommand{\lhrvcx}{\widehat{X}^n}
\newcommand{\lhrvcxone}{\widehat{X}_1^n}
\newcommand{\lhrvcxtwo}{\widehat{X}_2^n}
\newcommand{\lcrvcx}{\check{X}^n}
\newcommand{\lcrvcxone}{\check{X}_1^n}
\newcommand{\lcrvcxtwo}{\check{X}_2^n}
\newcommand{\rvcci}{{C}_i^n}
\newcommand{\rvcki}{{K}_i^n}
\newcommand{\rvcxi}{{X}_i^n}
\newcommand{\rvcyi}{{Y}_i^n}
\newcommand{\hrvcxi}{\widehat{X}_i^n}
\newcommand{\crvcxi}{\check{X}_i^n}
\newcommand{\vcki}{{k}_i^n}
\newcommand{\vcsi}{{s}_i^n}
\newcommand{\vcti}{{t}_i^n}
\newcommand{\vcvi}{{v}_i^n}
\newcommand{\vcwi}{{w}_i^n}
\newcommand{\vcxi}{{x}_i^n}
\newcommand{\vcyi}{{y}_i^n}

\newcommand{\vcs}{{s}^n}
\newcommand{\vct}{{t}^n}
\newcommand{\vcv}{{v}^n}
\newcommand{\vcw}{{w}^n}
%
%

\newcommand{\SZZ}{

\newcommand{\vcc}{{\vc c}}
\newcommand{\vck}{{\vc k}}
\newcommand{\vcx}{{\vc x}}
\newcommand{\vcy}{{\vc y}}
\newcommand{\vcz}{{\vc z}}
\newcommand{\vckone}{{\vc k}_1}
\newcommand{\vcktwo}{{\vc k}_2}
\newcommand{\vcxone}{{\vc x}_1}
\newcommand{\vcxtwo}{{\vc x}_2}
\newcommand{\vcyone}{{\vc y}_1}
\newcommand{\vcytwo}{{\vc y}_2}

\newcommand{\cvcx}{\check{\vc x}}
\newcommand{\cvcy}{\check{\vc y}}
\newcommand{\cvcz}{\check{\vc z}}
\newcommand{\cvcxone}{\check{\vc x}_1}
\newcommand{\cvcxtwo}{\check{\vc x}_2}

\newcommand{\hvcx}{\widehat{\vc x}}
\newcommand{\hvcy}{\widehat{\vc y}}
\newcommand{\hvcz}{\widehat{\vc z}}
\newcommand{\hvckone}{\widehat{\vc k}_1}
\newcommand{\hvcktwo}{\widehat{\vc k}_2}
\newcommand{\hvcxone}{\widehat{\vc x}_1}
\newcommand{\hvcxtwo}{\widehat{\vc x}_2}

\newcommand{\lvcc}{{c}}
\newcommand{\lvck}{{k}}
\newcommand{\lvcx}{{x}}
\newcommand{\lvcy}{{y}}
\newcommand{\lvcz}{{z}}

\newcommand{\lvckone}{{k}_1}
\newcommand{\lvcktwo}{{k}_2}

\newcommand{\lvcxone}{{x}}

\newcommand{\lvcxtwo}{{x}_2}
\newcommand{\lvcyone}{{y}_1}
\newcommand{\lvcytwo}{{y}_2}

\newcommand{\clvcxone}{\check{x}_1}
\newcommand{\clvcxtwo}{\check{x}_2}

\newcommand{\hlvckone}{\widehat{k}_1}
\newcommand{\hlvcktwo}{\widehat{k}_2}
\newcommand{\hlvcxone}{\widehat{x}_1}
\newcommand{\hlvcxtwo}{\widehat{x}_2}

\newcommand{\rvcc}{{\vc C}}
\newcommand{\rvck}{{\vc K}}
\newcommand{\rvcx}{{\vc X}}
\newcommand{\rvcy}{{\vc Y}}
\newcommand{\rvcz}{{\vc Z}}
\newcommand{\rvccone}{{\vc C}_1}
\newcommand{\rvcctwo}{{\vc C}_2}
\newcommand{\rvckone}{{\vc K}_1}
\newcommand{\rvcktwo}{{\vc K}_2}

\newcommand{\rvcxone}{{\vc X}}

\newcommand{\rvcxtwo}{{\vc X}_2}
\newcommand{\rvcyone}{{\vc Y}_1}
\newcommand{\rvcytwo}{{\vc Y}_2}
\newcommand{\hrvcxone}{\widehat{\vc X}_1}
\newcommand{\hrvcxtwo}{\widehat{\vc X}_2}

\newcommand{\lrvcc}{{C}}
\newcommand{\lrvck}{{K}}
\newcommand{\lrvcx}{{X}}
\newcommand{\lrvcy}{{Y}}
\newcommand{\lrvcz}{{Z}}
\newcommand{\lrvckone}{{K}_1}
\newcommand{\lrvcktwo}{{K}_2}
\newcommand{\lrvcxone}{{X}_1}
\newcommand{\lrvcxtwo}{{X}_2}
\newcommand{\lrvcyone}{{Y}_1}
\newcommand{\lrvcytwo}{{Y}_2}
\newcommand{\rvcci}{{\vc C}_i}
\newcommand{\rvcki}{{\vc K}_i}
\newcommand{\rvcxi}{{\vc X}_i}
\newcommand{\rvcyi}{{\vc Y}_i}
\newcommand{\hrvcxi}{\widehat{\vc X}_i}
\newcommand{\vcki}{{\vc k}_i}
\newcommand{\vcsi}{{\vc s}_i}
\newcommand{\vcti}{{\vc t}_i}
\newcommand{\vcvi}{{\vc v}_i}
\newcommand{\vcwi}{{\vc w}_i}
\newcommand{\vcxi}{{\vc x}_i}
\newcommand{\vcyi}{{\vc y}_i}

\newcommand{\vcs}{{\vc s}}
\newcommand{\vct}{{\vc t}}
\newcommand{\vcv}{{\vc v}}
\newcommand{\vcw}{{\vc w}}
}

\newcommand{\loF}{\underline{F}}

\newcommand{\prmtA}{\mu}
\newcommand{\prmtB}{\bar{\mu}}

\newcommand{\OmitZZa}{
}{
\section{Introduction \label{sec:introduction}}

We study the universal coding problem for 
the general framework of source encryption 
under side-channel attacks. 
This framework was posed 
and investigated by Oohama 
and Santoso \cite{DBLP:conf/isit/OohamaS22}. 
In this work  they proposed a general framework
for analyzing \emph{any} source encryption
with a symmetric key under \emph{any} kind of
side-channel attacks from which the adversary 
obtains some leaked information on the secret key.

Santoso and Oohama \cite{santosoOh:19} investigated the similar problem. 
Their work is  limited only 
to a single \emph{specific} encryption scheme, i.e., one-time-pad encryption. On the other hand, the framework provided by 
Oohama and Santoso \cite{DBLP:conf/isit/OohamaS22} covers
\emph{any} encryption scheme.
They further proposed a new security criterion for secrecy
which is defined as 
\emph{the maximum of all conditional mutual information
	between the ciphertext and plaintext given
	the adversarial key leakage, taken over
	all probability distributions of plaintexts}.

The reliable and secure rate region indicating 
the trade off between the compression rate of the ciphertext
and the rate constraint imposed on the adversary for secure 
source transmission was established by Oohama and Santoso \cite{DBLP:conf/isit/OohamaS22}. In this paper we focus on our attention to strengthening the direct coding theorem. We prove the existence of encryption/decryption schemes,
which are universal in the sense that they work effectively  
for {\it any distributions of the plain text}, {\it any noisy 
channels through which the adversary observe the corrupted version
of the key}, and any measurement device used for collecting 
the physical information, providing another proof of the 
direct coding theorem for the reliability and security 
rate region. Those schemes have a good performance  such that if we 
compress the ciphertext 
with 
rate within the achievable rate region, then: (1) anyone with 
secret key will be able to decrypt and decode the ciphertext 
correctly, but (2) any adversary who obtains the ciphertext 
and also the side physical information will not be able to obtain 
any information about the hidden source 
as long as the leaked physical information is 
encoded with a rate within the rate constraint.

}


\newcommand{\DelOneA}{
\section{Introduction \label{sec:introduction}}
As more cryptographic devices are deployed in
open physical spaces,
a new security challenge has risen
in the form of attackers which launch
\emph{side-channel attacks}, where
an attacker does not only collect the encrypted
data sent to the public communication channel,
but also collect
physical information related to the secret data
which are leaked by the devices
such as power consumption,
electromagnetic radiation, running time, etc.
Therefore, 
\emph{how to design an encryption scheme
	which is guaranteed to be secure
	even under side-channel attacks}
is a very important issue.

In this paper, we propose a general framework
for analyzing \emph{any} source encryption
with a symmetric key under \emph{any} kind of
side-channel attacks where the attacker
obtains some leaked information on the secret key.
Although  Santoso and Oohama
investigated the similar problem in
\cite{santosoOh:19},
their work is  limited only to a single \emph{specific}
encryption scheme, i.e., one-time-pad encryption,
while our framework covers
\emph{any} encryption scheme.
Then, we propose a new security criterion for secrecy
which is defined as \emph{the maximum of all conditional mutual information
	between the ciphertext and plaintext given
	the adversarial key leakage, taken over
	all probability distributions of plaintexts}.
One can easily see that our new security criterion is
more strict than the standard security criterion, i.e.,
the ordinary mutual information, as it is a usual practice
to derive an upper-bound of the security criterion
in order to guarantee secrecy of an encryption scheme.
Nevertheless, we show in this paper that we can construct
a concrete encryption scheme with reliable decoding 
and secrecy under side-channel attacks on the secret key.


The most important result obtained from
the modeling of our framework and the new security criterion is  
the strong converse, i.e., the necessary condition  
to have an encryption scheme with both reliable decoding and secrecy under side-channel attacks on the secret key. At the heart of our proof of the strong converse is the measurement of correlation 
between the ciphertexts and the information obtained by 
the adversary via side-channel attacks, 
given distribution of the plaintexts.  
}

\section{Problem Formulation}

\subsection{Preliminaries}

In this subsection, we show the basic notations and 
related consensus used in this paper. 

\noindent{}
\textit{Random Source of Information and Key: \ }
Let $X$ and $K$ be two random variables from a finite set 
$\mathcal{X}$. The random variable $X$ 
represents the source. The random $K$ 
represents the key used for encryption.
Let $\{X_t\}_{t=1}^\infty$ be a stationary discrete
memoryless source (DMS) such that for each $t=1,2,\ldots$, 
$X_{t}$ takes values in finite set $\mathcal{X}$ 
and has the same distribution as that of $X$ denoted by 
${p}_{X}=\{{p}_{X} (x)\}_{x \in \mathcal{X}}$.
The stationary DMS $\{X_t\}_{t=1}^\infty$ 
is specified with ${p}_{X}$. Similarly, 
let $\{K_{t}\}_{t=1}^\infty$ be a stationary 
DMS specified with the distribution ${p}_{K}$ of $K$. 

\noindent{}\textit{Random Variables and Sequences: \ }
We write the sequence of random variables with length $n$ 
from the information source as follows:
${\rvcx}\defeq X_{1}X_{2}\cdots X_{n}$. 
Similarly, the strings with length $n$ of $\mathcal{X}^n$ 
are written as 
${\vcx}\defeq x_{1}x_{2}\cdots
x_{n}\in\mathcal{X}^n$. 
For ${\vcx}\in \mathcal{X}^n$, 
${p}_{{\lrvcx}}({\vcx})$ stands for the 
probability of the occurrence of  
${\vcx}$. 
In the case of DMS, we write ${p}_{{\lrvcx}}({\vcx})$
as ${p}_{X}^n({\vcx})$. Similar notations are used 
for other random variables and sequences.

\noindent{}\emph{Consensus and Notations: }
Without loss of generality, throughout this paper,
we assume that $\mathcal{X}$ is a finite field.
The notation $\oplus$ is used to denote the field 
addition operation, while the notation $\ominus$ 
is used to denote the field subtraction operation, i.e., 
$a\ominus b = a \oplus (-b)$ for any elements 
$a,b \in {\cal X}$. Throughout 
this paper all logarithms are 
taken to the base natural.

\subsection{Basic System Description}

In this subsection we explain the basic system setting and 
basic adversarial model we consider in this paper. The information
source and the key are generated independently by different
parties $\Sgen$ and $\Kgen$ respectively. The source is 
generated by $\Sgen$ and independent of the key. 

\noindent
\underline{\it Source coding without encryption:} \ The 
random source ${\rvcx}$ from $\Sgen$ be sent to node 
$\mathsf{E}$. Further settings of the system are 
described as follows. 
\newcommand{\dElOne}{
Those are also shown in Fig. \ref{fig:mainA}.
\begin{figure}[t]
\centering
\includegraphics[width=0.43\textwidth]{SCGnl.eps}
\caption{Source coding without encryption.
\label{fig:mainA}}
\end{figure}
}
\begin{enumerate}
	\item \emph{Encoding Process:} \ 
        At the node $\mathsf{E}$, the encoder function 
        $\phi^{(n)}: {\cal X}^n $ $\to {\cal X}^{m}$ 
        observes ${\rvcx}$ to generate 
        $\tilde{X}^{m}=\phi^{(n)}({\rvcx})$. 
        Without loss of generality we may assume that 
        $\phi^{(n)}$ is {\it surjective}. 
	\item \emph{Transmission:} \ 
        Next, the encoded source $\tilde{X}^{m}$ is 
        sent to the destination $\D$ through a \emph{noiseless} 
        channel. 
	\item \emph{Decoding Process:} \ 
        In $\D$, the decoder function observes $\tilde{X}^{m}$
        to output ${\hrvcx}$, 
        using the one-to-one mapping $\psi^{(n)}$ defined by 
        $\psi^{(n)}:
             {\cal X}^{m} \to {\cal X}^n$. 
        Here we set 
        \begin{align*}
         \hrvcx \defeq & \psi^{(n)}(\tilde{X}^{m})
         = \psi^{(n)}\left(\phi^{(n)}(\rvcx)\right).
        \end{align*}
\end{enumerate}
For the above $(\phi^{(n)},\psi^{(n)})$, 
define the set $ \mathcal{D}^{(n)}$ of correct decoding by 
$
\mathcal{D}^{(n)} :=
\{\vcx \in \mathcal{X}^{n}:\psi^{(n)}(\phi^{(n)}(\vcx))=\vcx\}.
$
On $|{\cal D}^{(n)}|$, we have the following property. 
\begin{property}\label{pr:prOnDecSet} 
Under the conditions 1)-3), 
$|{\cal D}^{(n)}|=|{\cal X}^{m}|$. 
\end{property}

\newcommand{\ASSxx}{
\begin{IEEEproof}
We have the following:
\begin{align}
{\cal D}^{(n)}\MEq{a}& 
\{\vcxone =\psi^{(n)}(\tilde{x}^{m}):
\tilde{x}^{m} \in \phi^{(n)}({\cal X}^{n})\}
\notag\\
\MEq{b}&
\{ \vcx=\psi^{(n)}(\tilde{x}^{m}):
\tilde{x}^{m} \in {\cal X}^{m} \}.
\label{eqn:pSddxcc} 
\end{align}
Step (a) follows from that every pair 
$\tilde{x}^{m} \in \phi^{(n)}({\cal X}^{n})$ uniquely 
determines $\vcx \in {\cal D}^{(n)}$.
Step (b) follows from that $\phi^{(n)}$ are surjective.
Since 
$\psi^{(n)}: {\cal X}^{m} \to {\cal X}^n$ is a one-to-one mapping 
and (\ref{eqn:pSddxcc}), 
we have $|{\cal D}^{(n)}|=|{\cal X}^{m}|.$
\end{IEEEproof}
}


\newcommand{\ProofPrOnDecSet}{
\subsection{
Proof of Property \ref{pr:prOnDecSet} 
}
\label{apd:ProofPrOnDecSet}
In this appendix we prove the property on 
the decoding set ${\cal D}^{(n)}$ stated 
in Property \ref{pr:prOnDecSet}. 

\begin{IEEEproof}[Proof of Property \ref{pr:prOnDecSet}] 
We have the following:
\begin{align}
{\cal D}^{(n)}\MEq{a}& 
\{\vcxone =\psi^{(n)}(\tilde{x}^{m}):
\tilde{x}^{m} \in \phi^{(n)}({\cal X}^{n})\}
\notag\\
\MEq{b}&
\{ \vcx=\psi^{(n)}(\tilde{x}^{m}):
\tilde{x}^{m} \in {\cal X}^{m} \}.
\label{eqn:pSddxcc} 
\end{align}
Step (a) follows from that every pair 
$ \tilde{x}^{m} \in \phi^{(n)}({\cal X}^{n})$ uniquely 
determines $\vcx \in {\cal D}^{(n)}$.
Step (b) follows from that $\phi^{(n)}$ are surjective.
Since $\psi^{(n)}: {\cal X}^{m} \to {\cal X}^n$ is a one-to-one mapping 
and (\ref{eqn:pSddxcc}), we have $|{\cal D}^{(n)}|=|{\cal X}^{m}|.$
\end{IEEEproof}
}
\begin{remark}
For any $\tilde{\phi}^{(n)}$ not necessary {\it surjective}, 
we can construct ${\phi}^{(n)}$ such that   
${\phi}^{(n)}$ is {\it surjective} and 
$\tilde{\phi}^{(n)}({\cal X}^n)={\phi}^{(n)}({\cal X}^n)$.
Hence the assumption that $\phi^{(n)}$ is {\it surjective} 
causes no loss of generality.
\end{remark}

\noindent
\underline{\it Source coding with encryption:} \ 
The source ${\rvcx}$ from $\Sgen$ is sent to the node 
$\mathsf{L}$. The random key ${\rvck}$ from $\Kgen$, 
is sent to $\mathsf{L}$. Further settings of our system 
are described as follows. 
Those are also shown in Fig. \ref{fig:mainZ}.
\begin{enumerate}
\item \emph{Source Processing:} At the node $\L$, ${\rvcx}$ 
is encrypted with the key ${\rvck}$ using the encryption function 
        $\Phi^{(n)}:{\cal X}^n \times {\cal X}^n$ 
       	$\to {\cal X}^{m}$. 
        The ciphertext $C^{m}$ of ${\rvcx}$ is given by 
        $C^{m}=\Phi^{(n)}({\rvck},{\rvcx})$. 
        \item \emph{Transmission:} \ Next, the ciphertext 
        $C^{m}$ is sent to the destination $\D$ through 
        the \emph{public} communication channel. 
        Meanwhile, the key ${\rvck}$ is sent 
        to $\D$ through the \emph{private} communication channel.
        \item \emph{Sink Node Processing:} \ In $\D$, we decrypt 
        the ciphertext $\hrvcx$ from $C^{m}$ using 
        the key ${\rvck}$ through the corresponding 
        decryption procedure $\Psi^{(n)}$ defined by 
        $ \Psi^{(n)}: {\cal X}^n \times {\cal X}^{m} 
         \to {\cal X}^n$.
        Here we set 
        $ \hrvcx \defeq \Psi^{(n)}({\rvck},C^{m}).$ 
\end{enumerate}

Fix any $ \rvck=\vck \in \mathcal{X}^{n}$.
For this $\rvck$ and 
for $(\Phi^{(n)},\Psi^{(n)})$,  
we define the set $ \mathcal{D}^{(n)}_{\lvck}$ 
of correct decoding by 
\begin{align*}
\mathcal{D}^{(n)}_{\lvck}
& := \{\vcx \in \mathcal{X}^{n}:
\Psi^{(n)}({\vck},\Phi^{(n)}(\vck,\vcx))=\vcx \}.
\end{align*}
We require that the cryptosystem 
$
(\Phi^{(n)},\Psi^{(n)})
$
must satisfy the following condition.

{\it Condition:} 
For each distributed source encryption system 
$(\Phi^{(n)},\Psi^{(n)})$,  
there exists a source coding system $(\phi^{(n)},\psi^{(n)})$ 
such that for any $\vck \in \mathcal{X}^{n}$ and for any 
$\vcx \in \mathcal{X}^{n}$, 
\begin{align*}
& \Psi^{(n)}(\vck,\Phi^{(n)}(\vck,\vcx)) 
  =\psi^{(n)}(\phi^{(n)}(\vcx)). 
\end{align*}
The above condition implies that 
${\cal D}^{(n)}={\cal D}^{(n)}_{\lvck},
  \forall \vck \in {\cal X}^n$.
\newcommand{\OmiTTzx}{
We have the following properties on ${\cal D}^{(n)}$. 
\begin{property}
\label{pr:prOne}
$\quad$
\begin{itemize}
\item[a)] 
If $\vcx, {\vcy} \in {\cal D}^{(n)}$ and 
   $\vcx \neq {\vcy}$, then 
$\Phi^{(n)}_{\lvck}(\vcx) \neq \Phi^{(n)}_{\lvck}({\vcy}).$
\item[b)] 
    $\forall \vck$ and $\forall c^{m}$, 
$\exists \vcx$ $\in {\cal D}^{(n)}$ such that  
$
\Phi^{(n)}_{\lvck}(\vcx)=c^{m}.
$
\end{itemize}
\end{property}

Proof of Property \ref{pr:prOne} is found 
in Appendix B 
of \cite{oohamaSa22SideChAtk}. 
}

\newcommand{\ProofPrOne}{
\subsection{
Proof of Property \ref{pr:prOne}
}\label{apd:ProofPrOne}

We first prove the part a) and next prove the part b).   
\begin{IEEEproof}[Proof of Property \ref{pr:prOne} part a)] 
Under $\vcx, \vcy \in {\cal D}^{(n)}$ and 
$\vcx \neq$  $\vcy$, we assume that 
\beq
 \Phi^{(n)}_{\lvck}({\vcx})=\Phi^{(n)}_{\lvck}({\vcy}).
\label{eqn:Assum} 
\eeq
Then we have the following 
\begin{align}
&\vcx \MEq{a}
\psi^{(n)}(\phi^{(n)}(\vck) \MEq{b} \Psi^{(n)}_{\lvck}
(\Phi^{(n)}_{\lvck}({\vcx}))  
\notag\\
& \MEq{c}
\Psi^{(n)}_{\lvck}(\Phi^{(n)}_{\lvck}(\vcy))
\MEq{d}
 \psi^{(n)}(\phi^{(n)}(\vcy))\MEq{e}{\vcy}.
\label{eqn:SdCCv}
\end{align}
Steps (a) and (e) follow from the definition of 
${\cal D}^{(n)}$. Step (c) follows from (\ref{eqn:Assum}).
Steps (b) and (d) follow from the relationship between 
$(\phi^{(n)},\psi^{(n)})$ and 
$(\Phi^{(n)}_{\lvck}, \Psi^{(n)}_{\lvck}).$
The equality (\ref{eqn:SdCCv}) contradics the first assumption.
Hence we must have Property \ref{pr:prOne} part a).
\end{IEEEproof}
\begin{IEEEproof}[Proof of Property \ref{pr:prOne} part b)] 
We assume that $\exists \vck$ and 
$\exists c^{m}$ such that $\forall \vcx \in {\cal D}^{(n)}$, 
$\Phi^{(n)}_{\lvck}(\vcx)$ $\neq$ $c^{m}$. Set 
$$
{\cal B} \defeq \left\{ \Phi^{(n)}_{\lvck}(\vcx): 
\vcx \in {\cal D}^{(n)}
\right\}.
$$  
Then by the above assumption we have
\begin{align}
&{\cal B} \subseteq {\cal X}^{m} -\left\{c^{m}\right\}.
\label{eqn:SdCCvpp}
\end{align} 
On the other hand we have 
\begin{align*}
   \Psi^{(n)}_{\lvck}({\cal B})
&= \left\{
   \Psi^{(n)}_{\lvck}( \Phi^{(n)}_{\lvck}(\vcx)):
     \vcx \in {\cal D}^{(n)}
\right\}
\\
&= \left\{
   \psi^{(n)}(\phi^{(n)}(\vcx)): \vcx \in {\cal D}^{(n)}
\right\}
={\cal D}^{(n)},
\end{align*}
which together with that $\Psi_{\lvc k}:$ ${\cal X}^{m}$ 
$\to$ ${\cal X}^{n}$
is a one-to-one mapping yields that 
\begin{align*}
& |{\cal B}|=|\Psi^{(n)}_{\lvck}({\cal B})|
 =|{\cal D}^{(n)}|=|{\cal X}^{m}|.
\end{align*} 
The above equality contradicts (\ref{eqn:SdCCvpp}). 
Hence we must have that $\forall \vck$, $\forall c^{m}$,
$\exists \vcx \in {\cal D}^{(n)}$ such that 
$\Phi^{(n)}_{\lvck}(\vcx)=c^{m}$.
\end{IEEEproof}
}

\newcommand{\Dell}{
}{
\begin{figure}[t]
\centering
\includegraphics[width=0.43\textwidth]{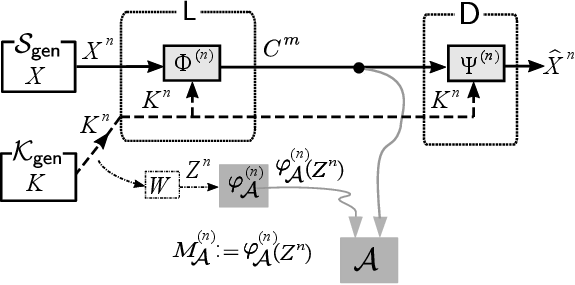}
\vspace*{-2mm}
\caption{Side-channel attacks to the source coding with encryption.
\label{fig:mainZ}}
\vspace*{-2mm}
\end{figure}}

\noindent
\underline{\it Side-Channel Attacks by Eavesdropper Adversary:} 
An 
\emph{adversary} $\A$ eavesdrops the public communication channel
in the system. The adversary $\A$ also uses a side information
obtained by side-channel attacks. 
Let ${\cal Z}$ be a finite set and let $W:{\cal X}\to {\cal Z}$ 
be a noisy channel. Let $Z$ be a channel output from $W$ for the 
input random variable $K$. Joint distribution $p_{KZ}$ of $(K,Z)$ is 
$p_{KZ}=(p_K,W)$. We consider the discrete memoryless 
channel specified with $W$. Let $\rvcz \in {\cal Z}^n$ be 
a random variable obtained as the channel output by connecting 
$\rvck \in {\cal X}^n$ to the input of channel. We write a 
conditional distribution on $\rvcz$ given $\rvck$ as 
$
W^n=
\left\{W^n(\vcz|\vck)\right
\}_{(\lvck,\lvcz)\in {\cal K}^n \times {\cal Z}^n}.
$
On the above output $\rvcz$ of $W^n$ for the input $\rvck$, 
we assume the followings.
\begin{itemize}
\item
The three random variables $X$, $K$ and $Z$, satisfy 
$X \perp (K,Z)$, which implies that $X^n \perp (K^n,Z^n)$.
\item $W$ is given in the system and the adversary ${\cal A}$ 
can not control $W$. 
\item By side-channel attacks, the adversary ${\cal A}$ 
can access $Z^n$. 
\end{itemize}
For each $n=1,2,\cdots$, 
let $\varphi_{\cal A}^{(n)}:{\cal Z}^n 
\to {\cal M}_{\cal A}^{(n)}$ be an encoder function. 
Set $M_{\cal A}^{(n)} \defeq \varphi_{\cal A}^{(n)}(Z^n) 
\in {\cal M}_{\cal A}^{(n)}$. We assume that 
$||\varphi_{\cal A}^{(n)}||=|{\cal M}_{\cal A}^{(n)}|$ must satisfy 
$|| \varphi_{\cal A}^{(n)}||\leq {\rm e}^{n R_{\cal A}}.$ 

\newcommand{\DeLoNe}{
Furthermore, set 
$\varphi_{\cal A} \defeq \{\varphi_{\cal A}^{(n)}\}_{n=1,2,\cdots}.$ 
Let 
$$ 
R_{\cal A}^{(n)}\defeq 
\frac{1}{n} \log ||\varphi_{\cal A}||
=\frac{1}{n} \log |{\cal M}_{\cal A}^{(n)}|
$$
be a rate of the encoder function $\varphi_{\cal A}^{(n)}$. 
In the following arguments all logarithms are 
taken to the base natural. 
For $R_{\cal A}>0$, we set 
${\cal F}_{\cal A}^{(n)}(R_{\cal A})\defeq 
\{ \varphi_{\cal A}^{(n)}: R_{\cal A}^{(n)} \leq R_{\cal A}\}.$ 
}

\subsection{
Security Criterion and Problem Set Up}

In this subsection we state our problem set up. 
This problem set up is due to Oohama and 
Santoso \cite{DBLP:conf/isit/OohamaS22}.
\newcommand{\OmiTTaz}{
In our previous works we introduced a security criterion which 
is a variant of the mutual information (the MI) criterion 
$I(C^{m}M_{\cal A}^{(n)};\rvcx) 
=I(C^{m};$ $\rvcx|M_{\cal A}^{(n)})$ 
$ 
\Delta_{\rm MI}^{(n)} 
\defeq I(C^{m}M_{\cal A}^{(n)};\rvcx)
=I(C^{m};$ $\rvcx|M_{\cal A}^{(n)})
$
between $\rvcx$ and $(C^{m},M_{\cal A}^{(n)})$. 
}
The adversary ${\cal A}$ tries to estimate 
${\rvcx} \in \mathcal{X}^n$ from 
$(C^{m},$ $M_{\cal A}^{(n)})$.
Note that since $X^n \perp (K^n,Z^n)$, we have 
$X^n \perp (K^n,$ $M_{\cal A}^{(n)})$. 
The mutual information (MI) between $\rvcx$ and 
$(C^{m},M_{\cal A}^{(n)})$ 
denoted by 
$
\Delta_{\rm MI}^{(n)} 
\defeq I(C^{m}M_{\cal A}^{(n)};\rvcx)
=I(C^{m};$ $\rvcx|M_{\cal A}^{(n)})
$
indicates a leakage of information on $\rvcx$ from 
$(C^{m},$ $M_{\cal A}^{(n)})$.  
Hence it seems to be quite natural 
to adopt the mutual information $\Delta_{\rm MI}^{(n)}$ 
as a security criterion. On the other hand, 
directly using $\Delta_{\rm MI}^{(n)}$ as a security criterion 
of the cyptosystem has some problem that this value 
depends on the statistical property of $\rvcx$. 
The following security criterion introduced by 
Oohama and Santoso \cite{DBLP:conf/isit/OohamaS22} is based on
$\Delta_{\rm MI}^{(n)}$ but not depending on the statistical 
property of $\rvcx$.  
\begin{definition} 
Let $\overline{X}^n $ be an arbitrary random variable 
taking values in ${\cal X}^n$. 
Set $\overline{C}^m=\Phi^{(n)}(\rvck,\overline{X}^n)$.
The maximum mutual information criterion denoted 
by $\Delta_{\rm max-MI}^{(n)}$ is as follows. 
\begin{align*}
&\Delta_{{\rm max-MI}}^{(n)} \defeq \max_{p_{\overline{X}^n} 
\in {\cal P}({\cal X}^n)} 
I(\overline{C}^{m} ; \overline{X}^n| M_{\cal A}^{(n)}).
\end{align*}
\end{definition} 

By definition it is obvious that 
$\Delta_{\rm MI}^{(n)} \leq {\Delta}_{{\rm max-MI}}^{(n)}$.
We have the following proposition on 
$\Delta_{{\rm max-MI}}^{(n)}$:
\begin{proposition}\label{pro:ProOneA}
$\quad$
\begin{itemize}
\item[a)] If we have $\Delta_{\rm MI}
=I(C^{m};\rvcx|M_{\cal A}^{(n)})=0,$ 
then, we have $\Delta_{{\rm max-MI}}^{(n)}=0$. 
This implies that $\Delta_{{\rm max-MI}}^{(n)}$ is valid as a 
measure of information leakage. 
\item[b)] $\:$
${\Delta}_{{\rm max-MI}}^{(n)} \geq m\log{|{\cal X}|}
 -H(K^{n}|M_{\cal A}^{(n)}).$
\end{itemize}
\end{proposition}

The property stated in the part b) is a key important property 
of $\Delta_{\rm max-MI}^{(n)}$, which plays an important role 
in establishing the strong converse theorem. 

\noindent
\underline{\it Defining Reliability and Security:} 
The decoding process is successful if $\hrvcx=\rvcx$ holds.
Hence the decoding error probability is given by  
\begin{align*}
&\Pr[\Psi^{(n)}(\rvck,\phi^{(n)}({\rvck, \rvcx}))\neq {\rvcxone}]
\\
&=\Pr[\psi^{(n)}(\phi^{(n)}(\rvcxone)) \neq \rvcxone]
=\Pr[\rvcxone \notin {\cal D}^{(n)}].
\end{align*}
Since the above quantity depends only on 
$(\phi^{(n)},\psi^{(n)})$ and ${p}_{X}^n$,
we write the error probability $p_{{\rm e}}$ of decoding as
\begin{align*}
p_{{\rm e}}=&p_{{\rm e}}(\phi^{(n)},\psi^{(n)}|{p}_{X}^n)
\defeq \Pr[\rvcxone \notin {\cal D}^{(n)}].
\end{align*}
We write $\Delta_{{\rm max-MI}}^{(n)}$ as 
$\Delta_{{\rm max-MI}}^{(n)}
(\varphi_{\cal A}^{(n)},\Phi^{(n)}|$
${p}_{KZ}^n)$ since it depends only on 
$\varphi_{\cal A}^{(n)}$, $\Phi^{(n)}$, and ${p}_{KZ}^n$.
Define  
\begin{align*}
&\Delta_{{\cal A}}^{(n)}(R_{\cal A},\Phi^{(n)}|{p}_{KZ}^n)
\\
&\defeq \max_{\varphi_{\cal A}^{(n)}}
\left\{
\Delta_{{\rm max-MI}}^{(n)}
(\varphi_{\cal A}^{(n)},\Phi^{(n)}|{p}_{KZ}^n):
||\varphi_{\cal A}^{(n)}||\leq {\rm e}^{nR_{\cal A}}
\right\}.
\end{align*}

\begin{definition} We fix some positive constant $\varepsilon_0$. 
        For a fixed pair $(\varepsilon, \delta) 
        \in [0,\varepsilon_0] \times (0,1)$, a quantity $R$ 
        is $(\varepsilon,\delta)$-admissible 
        under $R_{\cal A}$ $>0$
        for the system $\mathsf{Sys}$ if 
        $\exists \{(\Phi^{(n)},$ $\Psi^{(n)})\}_{n \geq 1}$
        such that $\forall \gamma >0$,
        $\exists n_0=n_0(\gamma) \in \mathbb{N}$, 
	$\forall n\geq n_0$, 
	\begin{align*}
        &(1/n)\log |{\cal X}^{m}| 
         = ({m}/{n})\log |{\cal X}| \in 
         \left[R-\gamma, R+ \gamma \right],
\\  				
& p_{{\rm e}}(\phi^{(n)},
              \psi^{(n)}|{p}_{X}^n)\leq \delta,\mbox{ and }
\Delta_{\cal A}^{(n)}(R_{\cal A},\Phi^{(n)}|p_{KZ}^n)
\leq \varepsilon.
\end{align*}
\end{definition}

\begin{definition}{\bf (Reliable and Secure Rate Region)}
        Let $\mathcal{R}_{\mathsf{Sys}}(\varepsilon,$ $\delta|
        {p}_{X},$ ${p}_{KZ})$
	denote the set of all $(R_{\cal A},R)$ such that
        $R$ is $(\varepsilon,\delta)$-admissible under $R_{\cal A}$. 
        Furthermore, set 
        $$
         \mathcal{R}_{\mathsf{Sys}}
          (p_{X},p_{KZ}) \defeq  
        \bigcap_{\scs (\varepsilon, \delta) \in (0,\varepsilon_0] 
                 {\scs \times (0,1)
                 }
         }
\mathcal{R}_{\mathsf{Sys}}(\varepsilon,\delta|p_{X},p_{KZ}),
$$
which we 
call the \emph{\bf reliable and secure rate} region. 
\end{definition}

Source encryption under side-channel attacks has a 
close connection with the privacy amplification for bounded storage 
eavesdropper posed and investigated by \cite{watanabe2012privacy}.
Oohama \cite{oohama:19} proved the exponential strong 
converse theorem for the one helper source coding problem 
posed and investigated by Ahlswede and K\"orner \cite{ahlswede:75} 
and Wyner \cite{wyner:75c}.   
Source encryption under side-channel attacks is also 
closely related to \cite{oohama:19}. 
Details of those relationships are described in \cite{DBLP:conf/isit/OohamaS22}. 

\section{Previous Results}

\subsection{
Strong  
Converse 
for Reliable and Secure Rate Region
}

In this subsection we state the results obtained by Oohama and Santoso \cite{DBLP:conf/isit/OohamaS22} on the characterization 
of the region 
$\mathcal{R}_{\mathsf{Sys}}(\varepsilon,$ $\delta| 
         {p}_{X},$ ${p}_{KZ})$. 
Let $U$ be an auxiliary random variable taking values in 
a finite set ${\cal U}$. We assume that the joint 
distribution $p_{UZK}$ of $(U,Z,K)$ has a form
$p_{U{Z}{K}}=(p_{UZ}, p_{K|Z})$. 
This condition is equivalent to $U \markov Z \leftrightarrow K$. 
Define the set of probability distribution $p=p_{UZK}$ by
\begin{align*}
&{\cal P}(p_{KZ})
\defeq 
\{p=p_{UZK}: \pa {\cal U} \pa 
\leq \pa {\cal Z} \pa+1, U \markov  Z \markov K \}.
\end{align*}
Let $\mathbb{R}_{+}^2:=\{ R_{\cal A}\geq 0,R\geq 0 \}$.
Let ${\cal R}_{\rm AKW}(p_{KZ})$ be the subset of 
$\mathbb{R}_{+}^2$ such that for some $U$ with 
$p_{UZK}\in {\cal P}(p_{KZ})$,
$
R_{\cal A}\geq I_{\empty}({Z};{U}), 
R \geq H_{\empty}({K}|{U}).
$ The region ${\cal R}_{\rm AKW}(p_{KZ})$ is equal to the rate region for the one helper source coding problem posed and investigated by Ahlswede and K\"orner \cite{ahlswede:75} 
and Wyner \cite{wyner:75c}. The subscript 
``AKW'' in ${\cal R}_{\rm AKW}(p_{KZ})$ is derived from their names. 
We can easily show that the region ${\cal R}_{\rm AKW}(p_{KZ})$ 
satisfies the 
following property.
\begin{property}\label{pr:pro0}  
$\quad$
\begin{itemize}
\item[a)] 
The region ${\cal R}_{\rm AKW}(p_{KZ})$ is a closed convex 
subset of $\mathbb{R}_{+}^2:=\{ R_{\cal A}\geq 0,R\geq 0 \}$.
\item[b)] The point $(0,H_{\empty}(K))$ always belongs to 
${\cal R}_{\rm AKW}(p_{KZ})$. Furthermore, for any $p_{KZ}$,
\begin{align*}
 {\cal R}_{\rm AKW}(p_{KZ}) \subseteq &
\{(R_{\cal A},R): R_{\cal A}+R \geq H_{\empty}(K)\} 
\cap \mathbb{R}_{+}^2.
\end{align*} 
\end{itemize}
\end{property}
\newcommand{\DelB}{
}{
\begin{figure}[t]
\centering
\includegraphics[width=0.32\textwidth]{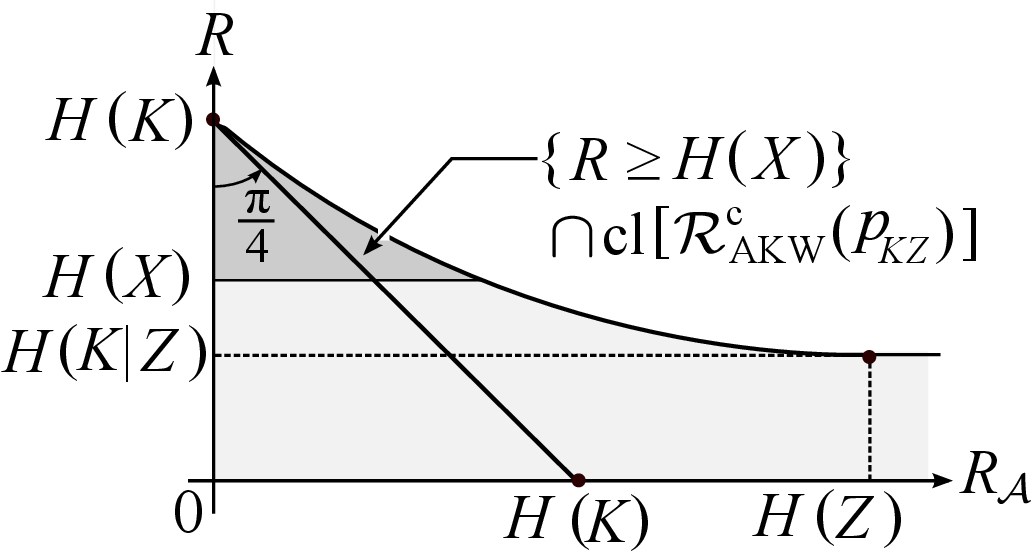}
\vspace*{-2mm}
\caption{Shape of the region 
${\cal R}(p_{X},p_{KZ})$.
}\label{fig:admissible}
\vspace*{-2mm}
\end{figure}
}

Set 
${\cal R}(p_{X},p_{KZ})
\defeq \{R \geq H(X)\} \cap {\rm cl}
[{\cal R}_{\rm AKW}^{\rm c}(p_{KZ})].
$
A typical shape of the region ${\cal R}(p_{X},p_{KZ})$ 
is shown in Fig. \ref{fig:admissible}. 
Then we have the following:
\begin{theorem}[Oohama and 
Santoso \cite{DBLP:conf/isit/OohamaS22}
]
\label{th:SStConvTh}

For each $(\varepsilon, \delta) 
\in (0,\varepsilon_0] \times (0,1)$, we have
\begin{align}
&\mathcal{R}_{\rm Sys}(p_{X},p_{KZ}) 
=
\mathcal{R}_{\rm Sys}(\varepsilon,\delta|
                       p_{X},p_{KZ})
={\cal R}(p_{X},p_{KZ}).
\end{align} 
\end{theorem}

The inclusion 
$\mathcal{R}_{\rm Sys}
(\varepsilon,\delta| p_{X},p_{KZ})
 \subseteq {\cal R}(p_{X},p_{KZ})$
can be proved by Property \ref{pro:ProOneA} part b), 
the direct coding theorem for one helper source coding, 
and a standard argument of the converse 
coding theorem. Details are found in 
   \cite{DBLP:conf/isit/OohamaS22}
In \cite{DBLP:conf/isit/OohamaS22}, 
Oohama and Santoso proved that 
if $(R_{\cal A},R) $ is an inner point of 
${\cal R}(p_{X},p_{KZ})$, then 
$\exists\{(\Phi^{(n)}, \Psi^{(n)})\}_{n=1}^{\infty}$ such that 
$ p_{{\rm e}}(\phi^{(n)},\psi^{(n)}
|{p}_{X}^n)$              
and $\Delta_{\cal A}^{(n)}(R_{\cal A}, \Phi^{(n)}|p_{KZ}^n)$
decay exponentially as $n$ tends to infinity, thereby 
establishing 
${\cal R}(p_{X},p_{KZ})\subseteq \mathcal{R}_{\rm Sys}
(p_{X},p_{KZ}).$
We state this result in the next subsection.  

\subsection{Direct Coding Theorem}

In this subsection we explain the direct coding theorem
established by Oohama and Santoso \cite{DBLP:conf/isit/OohamaS22}.  
We first define a function related to an exponential 
upper bound of $p_{\rm e}(\phi^{(n)},\psi^{(n)}|p_{X}^n)$.
Let $\overline{X}$ be an arbitrary random variable
over $\mathcal{X}$ and has a probability distribution 
$p_{\overline{X}}$. Let $\mathcal{P}(\mathcal{\cal X})$ 
denote the set of all probability distributions on 
$\mathcal{X}$. Fix $\gamma>0$, arbitrary. For $ R \geq 0$ and 
$p_{X} \in $ $\mathcal{P}(\mathcal{\cal X})$, 
we define the following function:
\begin{align*}
	E(R|p_{X}) &:{=}
	\min_{\scs p_{\overline{X}} \in 
	\mathcal{P}(\mathcal{\cal X}):
        \atop{\scs 
        R \leq H(\overline{X})}}
		D(p_{\overline{X}}||p_X).
\end{align*}
We next define a function related to an exponential 
upper bound of $\Delta_{\rm max-MI}^{(n)}
(\Phi^{(n)},\varphi_{\A}^{(n)}$ $|{p}_{KZ}^n)$.
Set
\begin{align*}
{\cal Q}(p_{K|{Z}}) \defeq& \{q=q_{U{Z}K}: 
\pa {\cal U} \pa \leq \pa {\cal {Z}} \pa,
{U} \markov {{Z}} \markov {K}, 
\\
& p_{K|{Z}}=q_{K|{Z}}\}.
\end{align*}
For $(\mu,\alpha) \in [0,1]^2$ and 
for $q=q_{U{Z}{K}}\in {\cal Q}(p_{K|{Z}})$, 
define 
\begin{align*}
& \omega_{q|p_{Z}}^{(\mu,\alpha)}({z},k|u)
 \defeq 
 \overline{\alpha} \log \frac{q_{{Z}}({z})}{p_{{Z}}({z})}
\\
&\quad + \alpha \left[
  {\mu}\log \frac{q_{{Z}|U}({z}|u)}{p_{{Z}}({z})}\right.
\left. +\overline{\mu}\log\frac{1}{q_{K|U}(k|u)}
\right],
\\
& \Omega^{(\mu,\alpha)}(q|p_{Z})
\defeq 
-\log {\rm E}_{q}
\left[\exp\left\{-\omega^{(\mu,\alpha)}_{q|p_{Z}}({Z},K|U)\right\}\right],
\\
& \Omega^{(\mu,\alpha)}(p_{KZ})
\defeq 
\min_{\scs 
   \atop{\scs 
    q \in {{\cal Q}}(p_{K|{Z}})
   }
}
\Omega^{(\mu,\alpha)}(q|p_{Z}).
\end{align*}
Furthermore, define 
\begin{align*}
& F(R_{\cal A}, R_{\empty}|p_{KZ})
\defeq \sup_{\scs (\mu,\alpha)
      \atop{\scs \in [0,1]^2
            }}
\frac{\Omega^{(\mu,\alpha)}(p_{KZ})
-\alpha({\mu}R_{\cal A} + \overline{\mu} R_{\empty})
}
{2+\alpha \overline{\mu}}.
\end{align*}

We can show that the above function satisfies 
the following:
\begin{property}\label{pr:pro1}  
$\quad$
\begin{itemize}
\item[a)] 
The cardinality bound 
$|{\cal U}|\leq |{\cal X}|$ in ${\cal Q}(p_{K|Z})$
is sufficient to describe the quantity
$\Omega^{({\mu,\alpha})}(p_{KZ})$. 
\item[b)] When $(R_{\cal A}+\tau, R +\tau) 
\notin {\cal R}(p_{KZ})$ for $\tau>0$, there 
exist $\lambda_0>0$ and $\mu_0 \in [0,1]$ such that
$$
F(R_{\cal A},R|p_{KZ}) > \frac{\tau}{2}
\cdot\frac{\lambda_0}{2+\lambda_0(5-\mu_0)}.
$$
\end{itemize}
\end{property}

Proofs the part a) of this property is found 
in Oohama \cite{oohama:19}. 
The part b) can be proved by 
some analytical computation. 
The detail of the proof is given 
in Appendix \ref{apd:ProofProOnePartC}.
\newcommand{\ProofProOnePartC}{
\subsection{
Proof of Property \ref{pr:pro1} part b)}
\label{apd:ProofProOnePartC}

Before stating the definition, we explain that 
the region ${\cal R}_{\rm AKW}(p_{KZ})$ can be expressed 
with a family of supporting hyperplanes. To describe 
this result, we define a set of probability distributions on 
${\cal U}$ $\times{\cal X}$ $\times{\cal Z}$ by
\begin{align*}
{\cal P}_{\rm sh}(p_{KZ})
&\defeq 
\{p=p_{UXY}: \pa {\cal U} \pa \leq \pa {\cal Z} \pa,
  U \markov  Z \markov K \}.
\end{align*}
For $\mu\in [0,1]$, define
\begin{align*}
& R^{(\mu)}(p_{KZ})
\defeq 
\min_{p \in {\cal P}_{\rm sh}(p_{KZ})}
\left\{{\mu} I(Z;U)+ \overline{\mu}H(K|U)\right\},
\end{align*}
where $\overline{\mu}=1-\mu$. Furthermore, define
\begin{align*}
& {\cal R}_{\rm AKW, sh}(p_{KZ})
\\
&\defeq \bigcap_{\mu \in [0,1]}\{(R_{\cal A},R_{\empty}):
\ba[t]{l}
{\mu} R_{\cal A}+ \overline{\mu}R_{\empty} \geq R^{(\mu)}(p_{KZ})\}.
\ea 
\end{align*}
Then, we have the following property.
\begin{property}\label{pr:pro0z} $\quad$
\begin{itemize}
\item[a)] The bound $|{\cal U}|\leq |{\cal X}|$
is sufficient to describe 
$R^{(\mu)}($ $p_{KZ})$. 
\item[b)] For any $p_{KZ},$ we have
\beq
 {\cal R}_{\rm AKW, sh}(p_{KZ})={\cal R}_{\rm AKW}(p_{KZ}).
\label{eqn:PropEqB}
\eeq
\end{itemize}
\end{property}

Proof this property is found in Oohama \cite{oohama:19}. 
We next define a function serving as a lower 
bound of $F(R_{\cal A}, R_{\empty}|p_{KZ})$. 
For $\lambda \geq 0$ and for  
$p_{UXY} \in {\cal P}_{\rm sh}(p_{KZ})$, define
\begin{align*}
& \tilde{\omega}_{p}^{(\mu)}({z},k|u)
\defeq {\mu}
  \log \frac{p_{{Z}|U}({z}|u)}{p_{{Z}}({z})}
  +\log \frac{1}{p_{K|U}(K|U)},
\\
& \tilde{\Omega}^{(\mu,\lambda)}(p)
\defeq 
-\log 
{\rm E}_{p}
\left[\exp\left\{-\lambda
\tilde{\omega}_p^{(\mu)}({Z},K|U)\right\}\right],
\\
& \tilde{\Omega}^{(\mu,\lambda)}(p_{KZ})
 \defeq \min_{\scs \atop{\scs 
p \in {{\cal P}_{\rm sh}(p_{KZ})}}}
\tilde{\Omega}^{(\mu,\lambda)}(p).
\end{align*}
Furthermore, define
\begin{align*}
&{\loF}(R_{\cal A},R|p_{KZ}) 
\\
& \defeq \sup_{ \lambda \geq 0,\mu \in [0,1]} 
\frac{\tilde{\Omega}^{(\mu,\lambda)}(p_{KZ})
-\lambda({\prmtA} R_{\cal A}+{\prmtB} R)}{2+\lambda(5-{\prmtA})}.
\end{align*}

We can show that the above functions
satisfy the following property.
\begin{property}\label{pr:pro1pp}  
$\quad$
\begin{itemize}
\item[a)] 
The cardinality bound 
$|{\cal U}|\leq |{\cal X}|$ in ${\cal P}_{\rm sh}(p_{KZ})$
is sufficient to describe the quantity
$\tilde{\Omega}^{(\mu,\lambda)}(p_{KZ})$. 

\item[b)] For any $R_{\cal A}, R \geq0$, we have 
\begin{align*}
& &F(R_{\cal A},R_{\empty}|p_{KZ})
\geq {\loF}(R_{\cal A},R_{\empty}|p_{KZ}).
\end{align*}
\end{itemize}
\end{property}

Proof this property is found in Oohama \cite{oohama:19}. 

{\it Proof of Property \ref{pr:pro1} part b):} 
By simple computation we have that for any $\mu \in [0,1]$ 
and any $p \in {\cal P}_{\rm sh}(p_{KZ})$, we have the 
following:  
\begin{align}
& \lim_{\lambda \to 0}
\frac{\tilde{\Omega}^{(\mu,\lambda)}(p)}{\lambda}
=\left(
\frac{\rm d}{ {\rm d} \lambda } 
\tilde{\Omega}^{(\mu,\lambda)}(p)
\right)_{\lambda=0}
\notag\\
&=\mu I(U;Z) + 
\overline{\mu}
H(K|U).
\label{eqn:QSdRRR}
\end{align} 
By the hyperplane expression ${\cal R}_{\rm AKW, sh}(p_{KZ})$ of 
${\cal R}_{\rm AKW}($ $p_{KZ})$ stated Property \ref{pr:pro0z} part b)
we have that when 
$(R_{\cal A}+\tau, R+\tau) \notin {\cal R}_{\rm AKW}(p_{KZ})$, 
we have 
\begin{align} 
&{\mu}_0 R_{\cal A} + \overline{\mu_0}R < R^{(\mu_0)}(p_{KZ})-\tau
\label{eqn:XddccPP}
\end{align} 
for some $\mu_0 \in [0,1]$. We fix $p \in {\cal P}_{\rm sh}(p_{KZ})$ 
arbitrary. By (\ref{eqn:QSdRRR}), 
there exists $\lambda_0>0$ such that 
\begin{align}
& \frac{\tilde{\Omega}^{(\mu_0,\lambda_0)}(p)}{\lambda_0}
\geq \mu_0 I(Z;U) + \overline{\mu_0} H(K|U)-\frac{\tau}{2}
\notag\\
& \geq R^{(\mu_0)}(p_{KZ})-\frac{\tau}{2}
\MGeq{a} {\mu}_0 R_{\cal A} + \overline{\mu_0}R +\frac{\tau}{2}.
\label{eqn:ZssP}
\end{align}
Step (a) follows from (\ref{eqn:XddccPP}). 
Since (\ref{eqn:ZssP}) holds for any 
$p \in {\cal P}_{\rm sh}(p_{KZ})$, we have
\beq 
{\tilde{\Omega}^{(\mu_0,\lambda_0)}(p_{KZ})}
\geq 
\lambda_0\left[{\mu}_0 R_{\cal A} + \overline{\mu_0}R
+\frac{\tau}{2}\right].
\label{eqn:ZsssF}
\eeq 
Then, we have the following chain of inequalities: 
\begin{align*}
&  F(R_{\cal A},R|p_{KZ}) \MGeq{a} \loF(R_{\cal A},R|p_{KZ}) 
\\
& \geq \frac{
\tilde{\Omega}^{(\mu_0,\lambda_0)}(p_{KZ})
-\lambda_0(\mu_0 R_{\cal A} + \overline{\mu_0} R)
      }
{1+\lambda_0(5-\mu_0)}
\\
& \MG{b} \frac{1}{2} \frac{\tau \lambda_0}
{1+\lambda_0(5-\mu_0)}.
\end{align*}
Step (a) follows from Property \ref{pr:pro1pp} part b).
Step (b) follows from (\ref{eqn:ZsssF}).
\hfill\IEEEQED
}
Oohama and Santoso \cite{DBLP:conf/isit/OohamaS22} obtained 
the following result. 
\begin{theorem}
[Oohama and Santoso 
\cite{DBLP:conf/isit/OohamaS22}]
\label{Th:mainth2}{
$\forall \gamma>0$, $\forall R_{\cal A}>0,$ $\forall R>0$, and 
$\forall p_{KZ}$ with $(R_{\cal A}, R)$ 
$\in {\cal R}_{\rm AKW}^{\rm c}(p_{KZ})$, 
$\exists n_0=n_0(\gamma)$ and 
$\exists \{(\Phi^{(n)},$ $\Psi^{(n)}) \}_{n \geq 1}$
satisfying 
$ (m/n) \log |{\cal X}|
\in \left[R-(1/n), R\right]$
such that $\forall n\geq n_0$ and 
$\forall p_X \in {\cal P}({\cal X})$,  
\begin{align}
& p_{\rm e}(\phi^{(n)},\psi^{(n)}|p_{X}^n) \leq 
(n+1)^{|{\cal X}|}{\ExP}^{-nE(R-\gamma|p_{X})}, 
\label{eqn:mainThErrB}\\
& \Delta_{\cal A}^{(n)}(R_{\cal A},\Phi^{(n)}|p_{KZ}^n)
\leq 5nR{\ExP}^{-nF(R_{\A},R|p_{KZ})}. 
\label{eqn:mainThSecB}
\end{align}
}\end{theorem}

We have the following corollary from 
Theorem \ref{Th:mainth2}. 
\begin{corollary}\label{cor:comainth2}{
$\forall R_{\cal A}>0,$ $\forall R>0$, and 
$\forall p_{KZ}$ with $(R_{\cal A}, R)$ 
$\in {\cal R}_{\rm AKW}^{\rm c}(p_{KZ})$, 
$\exists \{(\Phi^{(n)}, \Psi^{(n)}) \}_{n \geq 1}$
satisfying 
\begin{align*}
& \lim_{n \to \infty} (m/n)
\log |{\cal X}|=R,
\end{align*}
such that $\forall p_X$ with $R> H(X)$, 
\begin{align}
& \liminf_{n\to \infty}\left(-1/n\right)
\log p_{\rm e}(\phi^{(n)},\psi^{(n)}|p_{X}^n) 
\geq E(R|p_{X})>0, 
\label{eqn:mainThErrBab}
\\
& \liminf_{n\to \infty}\left(-1/n\right) 
   \log \Delta_{\cal A}^{(n)}(R_{\cal A},\Phi^{(n)} |p_{KZ}^n)
\notag\\
&\qquad \geq F(R_{\A},R|p_{KZ})>0. 
\label{eqn:mainThSecBab}
\end{align}
}\end{corollary}


By Corollary \ref{cor:comainth2}, under 
$(R_{\A},R)\in$  
$\{R>H(X)\}\cap {\cal R}_{\rm AKW}^{\rm c}($ $p_{KZ})$,
we have the followings: 
\begin{itemize}
\item[1.] On the reliability, 
$p_{\rm e}(\phi^{(n)},\psi^{(n)}|p_{X}^n)$ 
vanishes exponentially as $n\to\infty$, and its 
exponent is lower bounded by 
$E(R|p_{X})$.  
\item[2.] On the security, for any $\varphi_{\cal A}$ satisfying
$\varphi_{\cal A}^{(n)}\in$ ${\cal F}_{\cal A}^{(n)}(R_{\cal A})$, 
$\Delta^{(n)}(\varphi_{\A}^{(n)},\Phi^{(n)}$ 
$|p_{X}^n,p_{KZ}^n)$
vanishes exponentially as $n\to \infty$, and 
its exponent is lower bounded by $F(R_{\A},$ $R|p_{KZ})$.
\item[3.] The code that attains the exponent functions 
$E($$R|p_{X})$ is the universal code that depends
only on $R$ not on the value of the distribution $p_{X}$.
\end{itemize}

\begin{remark}
  The third result in the above items implies that the universality of 
 the code in Corollary \ref{cor:comainth2} is 
  {\it not perfect}. In fact, a construction of 
  the code attaining $F(R_{\A},$ $R|p_{KZ})$ depends 
  on $p_{KZ}=(p_{K}, W)$.      
\end{remark}

\section{Main Result}

In this paper we establish {\it a perfect universality} of codes attaining high reliability and security.      
To describe our main result we define several quantities.
For $p_{\overline{K}\,\overline{Z}} $ 
$\in {\cal P}({\cal X}\times {\cal Z})$, define 
\begin{align*}
& {G}(R_{\cal A},R_{\empty}|p_{\overline{K}\,\overline{Z}})
\defeq \sup_{\scs (\mu,\alpha)
\atop{\scs \in [0,1]^2}}
\frac{\Omega^{(\mu,\alpha)}(p_{\overline{K}\,\overline{Z}})
-\alpha({\mu}R_{\cal A} + \bar{\mu} R_{\empty})}
{2+3\alpha \bar{\mu}}.
%
%
\end{align*}
By simple computation we can show the following property.
\begin{property}\label{pr:PrExpG}
\begin{itemize}
\item[a)] $\forall \tau>0$, we have 
$$
{G}(R_{\cal A},R_{\empty}+\tau|p_{\overline{K}\,\overline{Z}})
\geq {G}(R_{\cal A},R_{\empty}|p_{\overline{K}\,\overline{Z}})
-(1/5)\tau.
$$
\item[b)]{$\quad$}
$
{G}(R_{\cal A},R_{\empty}|p_{\overline{K}\,\overline{Z}})
\geq (1/3){F}(R_{\cal A},R_{\empty}|p_{\overline{K}\,\overline{Z}}).
$
\end{itemize}
\end{property}

Set 
\begin{align*}
&
G(R_{\A},R|p_{KZ})
\\
&
\defeq 
\min_{
p_{\overline{K}\,\overline{Z}}
\in \mathcal{P}(\mathcal{\cal K} \times {\cal Z})}
\left\{ 
G(R_{\A},R|p_{\overline{K}\,\overline{Z}})
+D(p_{\overline{K}\,\overline{Z}}||p_{KZ}) \right\}.
\end{align*}
Our main result is the following. 
\begin{theorem}\label{Th:mainth3}
$\forall \gamma>0$, $\forall R_{\cal A}>0,$ and $\forall R>0$, 
$\exists n_0=n_0(\gamma)$ and
$\exists \{(\Phi^{(n)},$ $\Psi^{(n)})\}_{n \geq 1}$
satisfying 
$(m/n) \log |{\cal X}|
\in \left[R-(1/n), R\right]$
such that $\forall n\geq n_0$ and $\forall (p_X,$ $p_{KZ})$ 
$\in {\cal P}({\cal X})\times$ ${\cal P}({\cal X}\times {\cal Z})$,
\begin{align}
& p_{\rm e}(\phi^{(n)},\psi^{(n)}|p_{X}^n) \leq 
(n+1)^{|{\cal X}|}{\ExP}^{-nE(R-\gamma|p_{X})},
\label{eqn:mainThErrBb}\\
&  \Delta_{\cal A}^{(n)}(R_{\cal A},\Phi^{(n)}|p_{KZ}^n)
\notag\\
&\leq(7nR)(n+1)^{(16/5)|{\cal X}||{\cal Z}|}
      {\ExP}^{-nG(R_{\A},R|p_{KZ})}. 
\label{eqn:mainThSecBb}
\end{align}
\end{theorem}

We have the following corollary from Theorem \ref{Th:mainth3}. 
\begin{corollary}\label{cor:comainth3}{
$\forall R_{\cal A}>0$ and $\forall R>0$, 
$\exists \{(\Phi^{(n)}, \Psi^{(n)}) \}_{n \geq 1}$
satisfying 
\begin{align*}
& \lim_{n \to \infty} (m/n)\log |{\cal X}|=R,
\end{align*}
such that $\forall (p_X, p_{KZ})$ with 
$(R_{\A},R) \in$
$ \{ R>H(X)\}\cap {\cal R}_{\rm AKW}^{\rm c}$ $(p_{KZ})$,
\begin{align}
&\liminf_{n\to \infty}\left(-1/n\right)
\log p_{\rm e}(\phi^{(n)},\psi^{(n)}|p_{X}^n) 
\geq E(R|p_{X})>0, 
\label{eqn:mainThErrBc}\\
&\liminf_{n\to \infty}(-1/n) 
   \log \Delta_{\cal A}^{(n)}(R_{\cal A},\Phi^{(n)}|p_{KZ}^n)
\notag\\
& \qquad\qquad \geq G(R_{\A},R|p_{KZ})>0. 
\label{eqn:mainThSecBc}
\end{align}
}\end{corollary}

By Corollary \ref{cor:comainth3}, under $(R_{\A},R)\in$  
$\{R>H(X)\}\cap {\cal R}_{\rm AKW}^{\rm c}($ $p_{KZ})$,
we have the followings: 
\begin{itemize}
\item[1.] On the reliability, 
$p_{\rm e}(\phi^{(n)},\psi^{(n)}|p_{X}^n)$ 
goes to zero exponentially as $n$ tends to infinity, and its 
exponent is lower bounded by the function $E(R|p_{X})$.  
\item[2.] On the security, for any $\varphi_{\cal A}$ satisfying
$\varphi_{\cal A}^{(n)}\in$ 
${\cal F}_{\cal A}^{(n)}(R_{\cal A})$, 
$\Delta^{(n)}(\varphi_{\A}^{(n)},\Phi^{(n)}$ $|p_{KZ}^n)$
goes to zero exponentially as $n$ tends to infinity, and 
its exponent is lower bounded by the function 
$G(R_{\A},$$R|p_{KZ})$.
\item[3.] The code that attains the exponent functions 
$E($$R|p_{X})$ and $G(R_{\A},R|p_{K},W)$ is the universal 
code that depends only on $(R_{\cal A},R)$ not on the value 
of the distributions $p_{X}$ and $p_{KZ}=(p_K,$ $W)$.
\end{itemize}

\begin{remark}
  The third result in the above items implies that the universality of the code in Corollary \ref{cor:comainth3} is 
  {\it perfect}! 
\end{remark}

\section{Proof of Theorem \ref{Th:mainth3}}
In this section we prove Theorem \ref{Th:mainth3}.
We propose a new method to prove an existence of 
universal codes. Our method is a combination of the 
information spectrum method introduced 
by Han \cite{han:book} and 
the method of type developed 
by Csisz\'ar and K\"orner \cite{csi2011information}. 
\subsection{Types of Sequences and Their Properties 
}
For any $n$-sequence $k^n=k_{1}k_{2}\cdots $ 
$k_{n} \in {\cal X}^n$, 
$n(k|k^n)$ denotes the number of $t$ such that $k_{t}=k$.  
The relative frequency 
$\left\{n(k|k^n)/n\right\}_{k \in {\cal X}}$ 
of the components of ${\vcxone}$ is called the type of $k^n$ 
denoted by $P_{k^n}$. The set that consists of all 
the types on ${\cal X}$ is denoted by ${\cal P}_{n}({\cal X})$. 
For $p_{\overline{K}}\in {\cal P}_{n}({\cal X})$, let 
$T_{\overline{K}}^n$ denote the set of all $k^n$ such that 
$P_{k^n}=p_{\overline{K}}$.
Similarly, for any two $n$-sequences 
$k^n =k_{1}$ $k_{2}$ $\cdots$ $k_{n}\in $ 
${\cal X}^{n}$
and 
$z^n =z_{1}$ $z_{2}$ $\cdots$ $z_{n}\in $ 
${\cal Z}^{n}$, $n(k,z|k^n,z^n)$ 
denotes the number of $t$ such that $(k_{t},$ $z_{t})=(k,$ $z)$.
The relative frequency 
$
\{n(k,z|k^n,z^n)/n$ $\}_{
(k,z)\in}$ ${}_{ {\cal X} \times {\cal Z} }
$ of the components of $(k^n,z^n)$ 
is called the joint type 
of $(k^n,z^n)$ 
denoted by $P_{k^n,z^n}$. 
Furthermore, the set of all the joint type of 
${\cal X} \times {\cal Z}$ 
is denoted by ${\cal P}_n({\cal X} \times {\cal Z})$. 
For $p_{\overline{K}\,\overline{Z}}
\in {\cal P}_{n}({\cal X} \times {\cal Z})$, 
let
$T^n_{\overline{K}\,\overline{Z}}$ denote 
the set of all $(k^n,z^n)$ such that 
$P_{k^n, z^n}=p_{\overline{K}\,\overline{Z}}$.
Furthermore, for 
$p_{\overline{K}}\in$ ${\cal P}_{n}$ $({\cal X})$ and 
$k^n \in T_{\overline{K}}^n$, let  
$
T^n_{\overline{Z}|\overline{K} }(k^n)$
denote the set of all $z^n$ such that 
$P_{k^n,z^n}=p_{\overline{K}\,\overline{Z}}$.

The following lemma on the property of joint types are 
useful for proving several results related to 
the proof of Theorem \ref{Th:mainth3} and Theorem \ref{Th:mainth3}. 
For the detail of the proof see Csisz\'ar 
and K\"orner \cite{csi2011information}.
\begin{lemma}\label{lem:Lem1}{\rm $\quad$
\begin{itemize}
\item[a)]
$\begin{array}[t]{l} 
   |{\cal P}_{n}({\calVarX}\times\calVarZ)
   |\leq (n+1)^{|{\calVarX}||{\calVarZ}|}.
   \end{array}$
\item[b)] 
    For $p_{\ovKZ}\in {\cal P}_{n}({\calVarX \times \calVarZ})$,  
    the quantity ${\ExP}^{-nH({\ovKZ})}|T^n_{\ovKZ}|$ belongs to 
    the interval $[(n+1)^{-{|\calVarX||\calVarZ|}},1]$
\item[c)]
For any ${k^n}\in T_{\ovK}^n$, 
$|T^n_{\ovZ|\ovK}({k^n})|$ $={|T^n_{\ovKZ}|}/{|T^n_{\ovK}|}$.
\item[d)] For 
$(k^n,z^n) \in T^n_{\ovKZ}$,
\begin{align*}
&p_{KZ}^n(k^n,z^n)
={\ExP}^{-n[H(\ovKZ)+D(p_{\ovKZ}||p_{KZ})]}.
\end{align*}
 \end{itemize}}
\end{lemma} 

By Lemma \ref{lem:Lem1} parts b) and d), we immediately 
obtain the following lemma: 
\begin{lemma}\label{lem:Lem1.4}{\rm $\quad$
For $p_{\ovKZ}\in {\cal P}_{n}({\calVarX\times\calVarZ})$,  
\begin{align*}
& p_{KZ}^n(T_{\ovKZ}^n){\ExP}^{nD(p_{\ovKZ}||p_{KZ})}
\in [(n+1)^{-{|\calVarX||\calVarZ|}},1].
\end{align*}}
\end{lemma}

\subsection{A New Key Proposition}

For $p_{\overline{K}\,\overline{Z}}
\in {\cal P}_n({\cal X}\times {\cal Z})$, define 
\begin{align*}
&\Upsilon(\varphi_{\cal A}^{(n)},R|
 p_{\overline{K}\,\overline{Z}}) 
:=\sum_{\scs (a, k^n) 
  \atop{\scs \in {\cal M}_{\cal A}^{(n)} \times 
T^n_{\overline{K}} } }
\ts \frac{\left|\left(\varphi_{\cal A}^{(n)}\right)^{-1}(a)
\cap T^n_{\overline{Z}|\overline{K} }(k^n)\right|}
{\left|T^n_{\overline{K}\,\overline{Z}}\right|} 
\\
&\quad \times \log \left[1+({\rm e}^{nR}-1) 
\ts \frac{\left|\left(\varphi_{\cal A}^{(n)}\right)^{-1}(a)
\cap T^n_{\overline{Z}|\overline{K} }(k^n)\right|
\left| T^n_{\overline{Z}}\right|}
{\left|\left(\varphi_{\cal A}^{(n)}\right)^{-1}(a)
       \cap T^n_{\overline{Z}}\right|
\left|T^n_{\overline{K}\,\overline{Z}}\right|} 
\right].
\end{align*}
Then we have the following proposition.  
\begin{proposition}
\label{pro:UnivCodeBoundA} 
For any $R_{\cal A}, R>0$, there exists a sequence of mappings 
$\{(\Phi^{(n)}, \Psi^{(n)}) \}_{n=1}^{\infty}$
satisfying 
$({m}/{n})\log |{\cal X}| \in \left[R-(1/n), R \right]$
such that $\forall p_{KZ}$ 
$\in {\cal P}({\cal X} \times{\cal Z})$ and
$\forall \A$ with $\varphi_{\A}$ satisfying
$\varphi_{\A}^{(n)} \in {\cal F}_{\A}^{(n)}(R_{\A})$, we have
\begin{align}
& 
\Delta_{\rm max-MI}^{(n)}(\varphi_{\A}^{(n)},\Phi^{(n)}|p_{KZ}^n)
\leq (n+1)^{|{\cal X}||{\cal Z}|}
\notag\\
&\quad \times \hspace*{-1mm}
\sum_{\scs p_{\overline{K}\,\overline{Z} }
\atop{\scs \in {\cal P}_n( {\cal X} \times {\cal Z})}}
\hspace*{-2mm}
p_{KZ}^n(T_{\ovKZ}^n)
\Upsilon(\varphi_{\cal A}^{(n)},R|
p_{ \overline{K}\,\overline{Z} }). 
\label{eqn:mainThSecBnd}
\end{align}
\end{proposition}

Proposition \ref{pro:UnivCodeBoundA} can be proved 
by a standard argument on random coding, e.g. \cite{hayashi:10} 
and a well known method to prove an existence of universal codes 
via random coding argument, e.g. XIII in \cite{hayashi_matsu:16}. 
The detail of the proof is given in Appendix \ref{apd:PrProTwo}.  
The following is {\it a new key important result} for 
proving Theorem \ref{Th:mainth3}.   
\begin{proposition}\label{pro:KeyPro}
$\forall p_{\overline{K}\,\overline{Z}} 
\in {\cal P}_n({\cal X} \times {\cal Z})$ and 
$\forall \A$ with $\varphi_{\A}$ satisfying
$\varphi_{\A}^{(n)} \in {\cal F}_{\A}^{(n)}(R_{\A})$,
\begin{align}
& \Upsilon(\varphi_{\cal A}^{(n)},R|
p_{ \overline{K}\,\overline{Z} })
\notag\\
& \leq   
 (7nR)(n+1)^{(6/5)|{\cal X}||{\cal Z}|}
      {\ExP}^{-nG(R_{\A},R|p_{\overline{K}\,\overline{Z}})}.
\label{eqn:Zxx}      
\end{align}
\end{proposition}

Proof of Proposition \ref{pro:KeyPro} is the most difficult 
part of arguments to derive the lower bound of universally 
attainable secure exponent function for the information leakage.    
This proposition can be proved by a new method, 
which is a  coupling of the information spectrum method
introduced by Han \cite{han:book} and
the method of types developed 
by Csisz\'ar and K\"orner \cite{csi2011information}. Concretely, we establish {\it a new information spectrum meta converse lemma} in which the combinatorial 
type counting argument is included. Using this lemma and  
the recursive method developed by Oohama \cite{oohama:19}, we can derive exponential strong converse bounds. The detail of the proof 
is given in Appendix \ref{apd:PrProThr}.

In the remaining part of this 
subsection, we give the proof of Theorem \ref{Th:mainth3}. 

{\it Proof of Theorem \ref{Th:mainth3}:}
The bound (\ref{eqn:mainThErrBb}) 
in Theorem \ref{Th:mainth3} 
has already be proved by Oohama and Santoso \cite{DBLP:conf/isit/OohamaS22}. 
This result is stated as Proposition \ref{pro:ProCoding} 
in Appendix \ref{apd:EncDecSchme}.
It suffices to prove the bound (\ref{eqn:mainThSecBb}) 
in this theorem. From Propositions \ref{pro:UnivCodeBoundA} 
and \ref{pro:KeyPro}, we have that 
for any $R_{\cal A}, R>0$, there exists a sequence of mappings 
$\{(\Phi^{(n)}, \Psi^{(n)}) \}_{n=1}^{\infty}$
satisfying 
$ ({m}/{n}) \log |{\cal X}|
\in \left[R-(1/n), R \right]$
such that $\forall p_{KZ}$ 
$\in {\cal P}({\cal X} \times{\cal Z})$ and
$\forall \A$ with $\varphi_{\A}$ satisfying
$\varphi_{\A}^{(n)} \in {\cal F}_{\A}^{(n)}(R_{\A})$, we have
\begin{align}
& \Delta_{\rm max-MI}^{(n)}(\varphi_{\A}^{(n)},\Phi^{(n)}|p_{KZ}^n)
  \leq  (7nR)(n+1)^{(11/5)|{\cal X}||{\cal Z}|}
 \notag\\
&\quad \times \hspace*{-1mm}
\sum_{\scs p_{\overline{K}\,\overline{Z} }
\atop{\scs \in {\cal P}_n( {\cal X} \times {\cal Z})}}
\hspace*{-2mm}
p_{KZ}^n(T_{\ovKZ}^n)
{\ExP}^{-nG(R_{\A},R|p_{\overline{K}\,\overline{Z}})}
\notag\\
& \MLeq{a} (7nR)(n+1)^{(11/5)|{\cal X}||{\cal Z}|}
\notag\\
& \quad \times 
\sum_{\scs p_{\overline{K}\,\overline{Z} }
\atop{\scs \in {\cal P}_n( {\cal X} \times {\cal Z})}}
\hspace*{-2mm}
{\ExP}^{-n[G(R_{\A},R|p_{\overline{K}\,\overline{Z}})+
D(p_{\ovKZ}||p_{KZ})]}
\notag\\
& \leq (7nR)(n+1)^{(11/5)|{\cal X}||{\cal Z}|} 
  |{\cal P}_n( {\cal X} \times {\cal Z})|
{\ExP}^{-nG(R_{\A},R|p_{{K}{Z}})} 
\notag\\
& \MLeq{b} 
 (7nR)(n+1)^{(16/5)|{\cal X}||{\cal Z}|} 
 {\ExP}^{-nG(R_{\A},R|p_{{K}{Z}})}.
 \label{eqn:SZXXvv}
\end{align}
Step (a) follows from Lemma \ref{lem:Lem1.4}.
Step (b) follows from Lemma \ref{lem:Lem1} part a).
Since we have the bound (\ref{eqn:SZXXvv}) for any 
$\varphi_{\A}^{(n)} \in {\cal F}_{\A}^{(n)}(R_{\A})$, 
we have the bound (\ref{eqn:mainThSecBb}) 
in Theorem \ref{Th:mainth3}.
\hfill\IEEEQED




\appendix
\ProofProOnePartC

\subsection{
Coding Scheme, Reliability and Security Analysis 
}\label{apd:EncDecSchme}

In this appendix we explain the coding scheme proposed 
by Oohama and Santoso \cite{DBLP:conf/isit/OohamaS22}.
In this paper we use their coding scheme. 

The coding scheme proposed by Oohama and Santoso \cite{DBLP:conf/isit/OohamaS22} is illustrated in Fig. \ref{fig:solution}. In this coding scheme the following rate constraint is assumed: 
\begin{align}
& (1/n)\log |{\cal X}^{m}|= (m/n) \log |{\cal X}|
\in \left[R-(1/n), R\right].
\label{eqn:RateCond}
\end{align}
Oohama and Santoso \cite{DBLP:conf/isit/OohamaS22}
adopted a well known universal code construction of 
$\{(\phi^{(n)},\psi^{(n)})\}_{n\geq 1}$, deriving 
an exponential upper bound of 
$p_{\rm e}(\phi^{(n)},\psi^{(n)}|p_{X}^n)$.  
They next provided a concrete construction of 
$\varphi^{(n)}$ in Fig. \ref{fig:solution}.
Based on  $(\phi^{(n)},\psi^{(n)})$ and 
$\varphi^{(n)}$, Oohama and Santoso constructed $(\Phi^{(n)},\Psi^{(n)})$ in Fig. \ref{fig:solution}. 
They further 
provided some preliminary observation on an upper 
bound of $\Delta_{{\rm max-MI}}^{(n)}$ 
$(\varphi_{\cal A}^{(n)},\Phi^{(n)} |{p}_{KZ}^n)$. 

\noindent 
\underline{\it Universal Code Construction of 
$\{(\phi^{(n)}, \psi^{(n)})\}_{n\geq 1}$:} Set $\delta_{n}$\\ $\defeq (1/n)\{|{\cal X}|\log(n+1)+1\}$.
Note that $\delta_{n} \to 0$ as $ n \to \infty$. 
Let $n_0=n_0(\gamma)$ be the minimum integer such that we 
have $\delta_{n} \leq \gamma$ for $n\geq n_0(\gamma)$. 
According to Oohama and Santoso \cite{DBLP:conf/isit/OohamaS22}, we 
have the following proposition.
\begin{proposition}\label{pro:ProCoding}
$\forall \gamma >0$, $\exists \{(\phi^{(n)}, 
                                 \psi^{(n)})\}_{n\geq 1}$ 
satisfying (\ref{eqn:RateCond}), such that 
$\forall p_X$ with $R > H(X)$ and 
$\forall n \geq n_0(\gamma)$, 
\begin{align}
& p_{\rm e}(\phi^{(n)},\psi^{(n)}|p_{X}^n) \leq 
(n+1)^{|{\cal X}|}{\ExP}^{-n E(R-\gamma|p_{X}) }.
\label{eqn:mainZxxB}
\end{align}
\end{proposition}

Proposition \ref{pro:ProCoding} stated as Theorem 3.19 
in Chapter 3 of \cite{HanKingo}, is a well known result on 
the universal coding for discrete memoryless sources. 
The proof is also found there. 


\noindent
\underline{\it Affine Encoder as Privacy Amplifier:} 
Let $A$ is a matrix with $n$ rows and $m$ columns. Entries of 
$A$ are from ${\cal X}$. Let $b^{m}\in \mathcal{X}^{m}$. 
Define the mapping 
$\varphi^{(n)}: {\cal X}^n \to {\cal X}^{m}$
by 
\begin{align}
\varphi^{(n)}({\vck}):=&{\vck} A \oplus b^{m} 
\quad \mbox{ for }{\vck} \in \mathcal{X}^n.
\label{eq:homomorphic}
\end{align}
The mapping $\varphi^{(n)}$ is called the 
affine mapping. 
\newcommand{\DelAAA}{
}{
\begin{figure}[t] 
\vspace*{2mm}
	\centering 
	\includegraphics[width=0.44\textwidth]
	{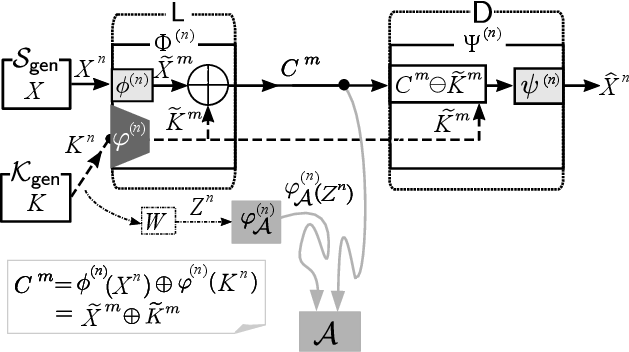}
\vspace*{-2mm}
	\caption{Proposed construction of 
        $(\Phi^{(n)},\Psi^{(n)})$.} 
        \label{fig:solution}%
\vspace*{-2mm}
\end{figure}
}

\noindent
\underline{\it Description of Proposed Procedure:} 
Construction of $(\Phi^{(n)},$ $\Psi^{(n)})$ 
is as follows: 
\begin{enumerate}
	\item \emph{Construction of $\Phi^{(n)}$:} \ 
        Define $\Phi^{(n)}: {\cal X}^{2n}\to {\cal X}^{m}$ by
        \begin{align*}
         \Phi^{(n)}(\vck,\vcx)
        &=\varphi^{(n)}(\vck)\oplus \phi^{(n)}(\vcx)
        \\ 
        &\quad \mbox{ for } \vck, \vcx\in {\cal X}^{n}.
        \end{align*}
        Let ${C}^{m}=\Phi^{(n)}(\rvcx,\rvck)$. 
        Send ${C}^{m}$ to the public communication 
        channel. Let $\wt{X}^{m}=\phi^{(n)}(\rvcx)$
        and $\wt{K}^{m}=\varphi^{(n)}(\rvck)$. Then we have  
${C}^{m}=\wt{X}^{m}\oplus \wt{K}^{m}.$
	\item\emph{Decoding at Sink Node $\D$: }
	First, using the linear encoder
	$\varphi^{(n)}$,
	$\D$ encodes the key $\rvck$ received
	through private channel into
	$\widetilde{K}^{m}=$$\varphi^{(n)}({\rvck})$.
	Receiving ${C}^{m}$ from
	public communication channel, $\D$ computes
	$\widetilde{X}^{m}$ in the following way.
        Since ${C}^{m}=\wt{X}^{m}\oplus \wt{K}^{m}$, 
        the decoder $\D$ can obtain 
        $\widetilde{X}^{m}$ $=\phi^{(n)}({\rvcx})$
        by subtracting $\widetilde{K}^{m}=\varphi^{(n)}({\rvck})$ 
        from ${C}^{m}$. 
	Finally, $\D$ outputs $\hrvcx$
	by applying the decoder 
        $\psi^{(n)}$ to 
	$\widetilde{X}^{m}$.
\end{enumerate}

\noindent
\underline{\it An Upper Bound of 
$\Delta_{{\rm max-MI}}^{(n)}( \varphi_{\cal A}^{(n)},\Phi^{(n)}
        |{p}_{KZ}^n)$:} 
We have the following upper bound of 
$\Delta_{{\rm max-MI}}^{(n)}(\varphi_{\A}^{(n)},\Phi^{(n)}|{p}_{KZ}^n)$. 
\begin{lemma}\label{lem:LemB} 
For the proposed construction of $\Phi^{(n)}$, 
we have 
$$
\Delta_{{\rm max-MI}}^{(n)}
(\varphi_{\cal A}^{(n)},\Phi^{(n)} |{p}_{KZ}^n)       
\leq m \log |{\cal X}|-H(\wt{K}^m| M_{\cal A}^{(n)}). 
$$
\end{lemma}
\begin{IEEEproof}
Let $\overline{X}^n $ be an arbitrary random variable 
taking values in ${\cal X}^n$. Set 
$\overline{C}^m=\Phi^{(n)}(\rvck,\overline{X}^n)$.
For the proposed construction of $\Phi^{(n)}$, we have 
$\overline{C}^m = \wt{K}^m \oplus \phi^{(n)}(\overline{X}^n).$
Then we have the following chain of inequalities:
\begin{align}
&  I(\overline{C}^{m}; \overline{X}^n| M_{\cal A}^{(n)})
=  H(\overline{C}^{m}|M_{\cal A}^{(n)})
  -H(\overline{C}^{m}|\overline{X}^n M_{\cal A}^{(n)})
\notag\\
&= H(\overline{C}^{m} | M_{\cal A}^{(n)})
  -H(\wt{K}^m \oplus \phi^{(n)}(\overline{X}^n)
|\overline{X}^n  M_{\cal A}^{(n)})
\notag\\
&\leq m \log |{\cal X}| - H(\wt{K}^m | 
\overline{X}^n M_{\cal A}^{(n)})
\notag\\
&
= m \log |{\cal X}|-H(\wt{K}^m | M_{\cal A}^{(n)}).
\label{eqn:AsDDss}
\end{align}
Since (\ref{eqn:AsDDss}) holds for any $\overline{X}^n$, we have
the upper bound of  
$
\Delta_{{\rm max-MI}}^{(n)}(\varphi_{\A}^{(n)},\Phi^{(n)}
|{p}_{KZ}^n)
$
in Lemma \ref{lem:LemB}. 
\end{IEEEproof}

\subsection{Proof of Proposition \ref{pro:UnivCodeBoundA}}
\label{apd:PrProTwo}

In this appendix we estimate upper bounds of 
$\Delta^{(n)}_{\rm max-MI}($ $\varphi_{\A}^{(n)},\Phi^{(n)}|p_{KZ}^n)$ to prove Proposition \ref{pro:UnivCodeBoundA}.
Set $M^{(n)} \defeq P_{K^n,Z^n}$.
Note that for $p_{\overline{K}\,\overline{Z}} 
\in {\cal P}_n({\cal  X}\times {\cal Z})$, 
$$
\Pr\{ M^{(n)}=p_{\overline{K}\,\overline{Z}}\}
= p_{KZ}^n(T_{\ovKZ}^n).
$$
For $p_{\overline{K}\,\overline{Z}} 
\in {\cal P}_n({\cal  X}\times {\cal Z})$, 
set 
\begin{align*}
& \zeta (\varphi_{\cal A}^{(n)},
\varphi^{(n)}|p_{\overline{K}\,\overline{Z}})
\notag\\
& \defeq 
\left.\left.\left.
D\left( p_{\widetilde{K}^m|M_{{\cal A}}^{(n)}M^{(n)}
       =p_{\overline{K}\,\overline{Z}}}
\right|\right| p_{V^m} \right| 
p_{M_{\cal A}^{(n)}M^{(n)}=p_{\overline{K}\,\overline{Z}}} \right),
\end{align*}
where $p_{V^m}$ is the uniform distribution over $\mathcal{X}^{m}$. 
Then we have the following:
\begin{lemma}\label{lem:LemC} 
\begin{align}
&\Delta^{(n)}_{\rm max-MI}(\varphi_{\A}^{(n)},\Phi^{(n)}|p_{KZ}^n)
\leq \sum_{\scs p_{\overline{K}\,\overline{Z} }
\atop{\scs \in {\cal P}_n( {\cal X} \times {\cal Z})}}1
\notag\\
&\quad \times 
p_{KZ}^n(T_{\ovKZ}^n)
\zeta (\varphi_{\cal A}^{(n)},\varphi^{(n)} 
|p_{\overline{K}\,\overline{Z}}).
\label{eqn:ZddFxx}
\end{align}
\end{lemma}

{\it Proof: } From Lemma \ref{lem:LemB}, 
we have the following:
\begin{align*}
&\Delta^{(n)}_{\rm max-MI}(\varphi_{\A}^{(n)},\Phi^{(n)}
|p_{KZ}^n) 
\notag\\
& \leq m\log |{\cal X}|-H(\widetilde{K}^m|
  M_{{\cal A}}^{(n)}M^{(n)})
\notag\\
&=\left.\left.\left.\!\!\!
D\left(p_{\widetilde{K}^m|M_{{\cal A}}^{(n)}M^{(n)}}
\right|\right|
p_{V^m} \right| p_{M_{\cal A}^{(n)}M^{(n)}}\right),
\end{align*}
from which and the definition of $\zeta (\varphi_{\cal A}^{(n)},
\varphi^{(n)}|p_{\overline{K}\,\overline{Z}})$, we have 
the bound (\ref{eqn:ZddFxx}) in Lemma \ref{lem:LemC}.
\hfill\IEEEQED

Fix any pair of random variables $(\overline{K},\overline{Z})$ such that
$p_{\overline{K}\,\overline{Z}} \in {\cal P}_n({\cal X}\times {\cal Z})$.
The following lemma is easily obtained by Lemma \ref{lem:Lem1} parts c) 
and d). 
\begin{lemma}\label{lem:lemFour}
$\forall k^n \in {\cal X}^n$, 
$\forall a\in {\cal M}_{\cal A}^{(n)}$,  
and $\forall p_{\overline{K}\,\overline{Z}}
\in {\cal P}_n({\cal X}\times {\cal Z})$, we have
\begin{align*}
& p_{K^n M_{\cal A}^{(n)} | M^{(n)} } 
(k^n,a| p_{\overline{K}\,\overline{Z}})
\\
&=\Pr\{K^n=k^n, M_{ {\cal A} }^{(n)}=a
| M^{(n)}=p_{\overline{K}\,\overline{Z}}\}
\\
&=\ts \frac{\left|\left(\varphi_{\cal A}^{(n)}\right)^{-1}(a)
\cap T^n_{\overline{Z}|\overline{K} }(k^n)\right|}
{\left|T^n_{\overline{K}\,\overline{Z}}\right|},
\\
& p_{ M_{\cal A}^{(n)}| M^{(n)}}
(a|p_{\overline{K}\,\overline{Z}})
= \Pr\{M_{\cal A}^{(n)}=a|M^{(n)}
=p_{\overline{K}\overline{Z}}\}
\\
&=\ts \frac{\left|\left(\varphi_{\cal A}^{(n)}\right)^{-1}(a)
\cap T^n_{\overline{Z}}\right|}
{\left|T^n_{\overline{Z}}\right|}.
\end{align*}
\end{lemma}

Using Lemma \ref{lem:lemFour}, $\Upsilon(\varphi_{\cal A}^{(n)},R|
 p_{\overline{K}\,\overline{Z}})$ can be written as 
\begin{align*}
&\Upsilon(\varphi_{\cal A}^{(n)},R|
 p_{\overline{K}\,\overline{Z}})
\\ 
& =\sum_{\scs (a, k^n) 
  \atop{\scs \in {\cal M}_{\cal A}^{(n)} \times 
T^n_{\overline{K}} } }
 p_{K^n M_{ {\cal A} }^{(n)} | M^{(n)} } 
(k^n,a| p_{\overline{K}\,\overline{Z}})
\\
& \quad \times \log \left[1+({\rm e}^{nR}-1) 
 p_{K^n| M_{ {\cal A} } ^{(n)}M^{(n)} } 
(k^n|a, p_{\overline{K}\,\overline{Z}})
\right].
\end{align*}
On $\zeta(\varphi_{\cal A}^{(n)},\varphi^{(n)}|
p_{\overline{K}\,\overline{Z}})$
for $p_{\overline{K}\,\overline{Z}}
\in {\cal P}_n({\cal X}\times {\cal Z})$, 
we have the following lemma.
\begin{lemma}\label{lem:RdBound}
$\forall p_{\overline{K}\,\overline{Z}} 
\in {\cal P}_n({\cal X} \times {\cal Z})$,
\begin{align}
&{\bf E}
\left[\zeta(\varphi_{\cal A}^{(n)},\varphi^{(n)}             
              | p_{\overline{K}\,\overline{Z}})
\right]
\leq \Upsilon(\varphi_{\cal A}^{(n)},R|
 p_{\overline{K}\,\overline{Z}}), 
\label{eqn:SDDx}
\end{align}
where ${\bf E}[\cdot]$ is an expectation based on 
the random choice of $\varphi^{(n)}$.
\end{lemma}

To prove the above lemma we use a technique quite similar 
to that of Hayashi \cite{hayashi:10} used for an ensemble 
of universal${}_2$ functions. 
Lemmas \ref{lem:LemC} and \ref{lem:RdBound} 
together with a standard argument of proving an 
existence of universal codes yield Proposition \ref{pro:UnivCodeBoundA}.   

{\it Proof of Proposition \ref{pro:UnivCodeBoundA}:} 
We have the following chain of inequalities:
\begin{align}
&{\bf E}
\left[\sum_{\scs p_{\overline{K}\,\overline{Z} }
\atop{\scs \in {\cal P}_n( {\cal X} \times {\cal Z})}}
\left( 
 \Upsilon(\varphi_{\cal A}^{(n)},R|
 p_{\overline{K}\,\overline{Z}})
 \right)^{-1}
\zeta(\varphi_{\cal A}^{(n)},\varphi^{(n)} 
  | p_{\overline{K}\,\overline{Z}})           
\right]
\notag\\
&=\sum_{\scs p_{\overline{K}\,\overline{Z} }
\atop{\scs \in {\cal P}_n( {\cal X} \times {\cal Z})}}
\left( 
 \Upsilon(\varphi_{\cal A}^{(n)},R|
 p_{\overline{K}\,\overline{Z}})
 \right)^{-1}{\bf E}\left[
\zeta(\varphi_{\cal A}^{(n)},\varphi^{(n)} 
  | p_{\overline{K}\,\overline{Z}})           
\right]
\notag \\
&\MLeq{a} \sum_{\scs p_{\overline{K}\,\overline{Z} }
\atop{\scs \in {\cal P}_n( {\cal X} \times {\cal Z})}}1
= |{\cal P}_n( {\cal X} \times {\cal Z})|
\MLeq{b} (n+1)^{|{\cal X}||{\cal Z}|}. 
\label{eqn:SDDxy}
\end{align}
Step (a) follows from Lemma \ref{lem:RdBound}.
Step (b) follows from Lemma \ref{lem:Lem1} part a).
From (\ref{eqn:SDDxy}), we can see that 
there exists at least one deterministic 
function $\varphi^{(n)}$ such that 
\begin{align}
& \sum_{\scs p_{\overline{K}\,\overline{Z} }
\atop{\scs \in {\cal P}_n( {\cal X} \times {\cal Z})}}
\left( 
 \Upsilon(\varphi_{\cal A}^{(n)},R|
 p_{\overline{K}\,\overline{Z}})
 \right)^{-1}
\zeta(\varphi_{\cal A}^{(n)},\varphi^{(n)} 
  | p_{\overline{K}\,\overline{Z}})  
\notag \\
&\leq  (n+1)^{|{\cal X}||{\cal Z}|}. 
\label{eqn:SDDxyz}
\end{align}
From (\ref{eqn:SDDxyz}), we have that 
$\forall p_{\ovKZ} \in 
{\cal P}_n( {\cal X} \times {\cal Z})$, 
\begin{align}
\zeta(\varphi_{\cal A}^{(n)},\varphi^{(n)} 
  | p_{\overline{K}\,\overline{Z}})
  \leq (n+1)^{|{\cal X}||{\cal Z}|}
   \Upsilon(\varphi_{\cal A}^{(n)},R|
 p_{\overline{K}\,\overline{Z}}).
 \label{eqn:SDDxyzx}
\end{align}
From (\ref{eqn:ZddFxx}) in Lemma \ref{lem:LemC} and (\ref{eqn:SDDxyzx}), we have the bound (\ref{eqn:mainThSecBnd}) 
in Proposition \ref{pro:UnivCodeBoundA}. 
\hfill\IEEEQED


\subsection{Proof of Proposition \ref{pro:KeyPro}
}\label{apd:PrProThr}

In this appendix we derive upper bounds of 
$\Upsilon(\varphi_{\cal A}^{(n)},$ $R|
p_{\overline{K}\,\overline{Z}})$ for 
$p_{\overline{K}\,\overline{Z}}$ 
$\in {\cal P}_n( {\cal X} \times {\cal Z})$ to 
prove Proposition \ref{pro:KeyPro}. 

Fix any $p_{\overline{K}\,\overline{Z}}$ 
$\in {\cal P}_n({\cal X} \times {\cal Z})$.
We assume that $(K^n,Z^n) \sim 
p^n_{\overline{K}\,\overline{Z}}$. 
Under this assumption, we define  
\begin{align}
& \Upsilon(\varphi_{\cal A}^{(n)},R|
   p^n_{\overline{K}\,\overline{Z}})
\notag\\
&\defeq \sum_{p_{\hat{K}\hat{Z}}
        \in {\cal P}_n({\cal X}\times{\cal Z})}
p_{\ovKZ}^n(T_{\hat{K}\hat{Z}}^n)
\Upsilon(\varphi_{\cal A}^{(n)},R|p_{\hat{K}\hat{Z}}).
\label{eqn:Saww}
\end{align}
Then we have the following lemma.
\begin{lemma}
\label{lem:Ohzzza}
For any $p_{\overline{K}\,\overline{Z}}$ 
$\in {\cal P}_n({\cal X} \times {\cal Z})$, we have
\begin{align}
\frac{\Upsilon(\varphi_{\cal A}^{(n)},R|
p_{\overline{K}\,\overline{Z}})}
{(n+1)^{|{\cal X}||{\cal Z}|}} 
\leq \Upsilon(\varphi_{\cal A}^{(n)},R|
p^n_{\overline{K}\,\overline{Z}}). 
\end{align}
\end{lemma}

{\it Proof:} By the definition (\ref{eqn:Saww}) of 
$\Upsilon(\varphi_{\cal A}^{(n)},R|
   p^n_{\overline{K}\,\overline{Z}})$, we have that
\begin{align}
& \Upsilon(\varphi_{\cal A}^{(n)},R|   p^n_{\overline{K}\,\overline{Z}})
\geq p_{\ovKZ}^n(T_{\ovKZ}^n)
\Upsilon(\varphi_{\cal A}^{(n)},R|p_{\ovKZ})
\notag\\
&\quad \MGeq{a}
(n+1)^{-|{\cal X}||{\cal Z}|}\Upsilon(\varphi_{\cal A}^{(n)},R|p_{\ovKZ}).
\label{eqn:Sawwxxxx}
\end{align}
Step (a) follows from Lemma \ref{lem:Lem1.4}.
\hfill\IEEEQED

To derive the upper bound of
$\Upsilon(\varphi_{\cal A}^{(n)},R|
   p^n_{\overline{K}\,\overline{Z}})$, 
we observe that
\begin{align}
&
\Upsilon(\varphi_{\cal A}^{(n)},R|
   p^n_{\overline{K}\,\overline{Z}})
={\rm E}\Bigl[
\log \Bigl\{1
\notag\\
&\quad+({\rm e}^{nR}-1) p_{K^n|M_{\cal A}^{(n)}M^{(n)}}
(K^n|M_{\cal A}^{(n)},M^{(n)})\Bigr\}\Bigr].
\label{eqn:AdttA}
\end{align}
Set
\begin{align}
&\wp_\eta^{(n)}
  = \wp^{(n)}_{\eta}(\varphi_{\cal A}^{(n)},R|p_{KZ}^n) \defeq p_{ M_{\cal A}^{(n)} Z^n K^n}\Biggl\{
  \nonumber\\
& R +\eta \geq \left.
 \frac{1}{n}{ \ts\log
 \frac{1}{p_{K^n|M_{\cal A}^{(n)}M^{(n)}}(K^n|M_{\cal A}^{(n)},M^{(n)})}}
 \right\}.
\end{align}
From (\ref{eqn:AdttA}), we have the following lemma.
\begin{lemma}\label{lem:LemDss}
$\forall \eta>0$, $\forall n \geq R^{-1}$, and 
$\forall\varphi_{\A}$ satisfying
$\varphi_{\A}^{(n)} \in {\cal F}_{\A}^{(n)}(R_{\A})$
\begin{align*}
&(nR)^{-1}\Upsilon(\varphi_{\cal A}^{(n)},R|
p^n_{\overline{K}\,\overline{Z}})
\leq \wp_\eta^{(n)}
(\varphi_{\cal A}^{(n)},R|
p^n_{\overline{K}\,\overline{Z}})+{\rm e}^{-n\eta}.
\end{align*}
\end{lemma}

Using Lemma \ref{lem:LemDss} and evaluating an upper bound of 
$\wp_\eta^{(n)}
($ $\varphi_{\cal A}^{(n)},R|
p^n_{\overline{K}\,\overline{Z}})$,
we have the following lemma:
\begin{lemma}
\label{lem:Ohzzz}
Fix any 
$p_{\overline{K}\,\overline{Z}}$ 
$\in {\cal P}_n({\cal X} \times {\cal Z})$.
Set $\tau_n\defeq (|{\cal X}||{\cal Z}|/n)\log (n+1)$.
Suppose that $(K^n,Z^n) \sim 
p^n_{\overline{K}\,\overline{Z}}$. 
Then, $\forall \eta>0$, $\forall n \geq R^{-1}$, and 
$\forall\varphi_{\A}$ satisfying
$\varphi_{\A}^{(n)} \in {\cal F}_{\A}^{(n)}(R_{\A})$,
we have
\begin{align}
& (nR)^{-1}\Upsilon(\varphi_{\cal A}^{(n)},R|
p^n_{\overline{K}\,\overline{Z}})
\notag\\
&
  \leq p_{M_{\cal A}^{(n)}M^{(n)}Z^nK^n}\Biggl\{
\eta  \geq \frac{1}{n} \log \ts \frac{ q_{{Z}^n}(Z^n)}{p_{Z^n}(Z^n)},
\notag\\
& 
\eta \geq \frac{1}{n}\log \ts 
\frac{\hat{q}_{M_{\cal A}^{(n)}Z^nK^n}(M_{\cal A}^{(n)},Z^n,K^n)}
{p_{M_{\cal A}^{(n)}Z^nK^n}(M_{\cal A}^{(n)},Z^n,K^n)},
\notag\\
& \eta \geq  \frac{1}{n} \log \ts
\frac{p_{Z^n|K^n M_{\cal A}^{(n)}}
        (Z^n|K^n, M_{\cal A}^{(n)})}
     {p_{Z^n|K^n M_{\cal A}^{(n)} M^{(n)}}
        (Z^n|K^n,M_{\cal A}^{(n)},M^{(n)})},
\label{eqn:NewboundOne} \\
&\tau_n +\eta \geq  \frac{1}{n}\log \ts 
\frac{p_{K^n Z^n| M_{\cal A}^{(n)}M^{(n)}}
        (K^n,Z^n| M_{\cal A}^{(n)},M^{(n)})}
     {p_{K^n Z^n| M_{\cal A}^{(n)}}
        (K^n,Z^n| M_{\cal A}^{(n)})},
\label{eqn:NewboundTwo} \\
& R_{\cal A}+\eta\geq  \frac{1}{n}\log \ts
\frac{p_{Z^n|M_{\cal A}^{(n)}}(Z^n|M_{\cal A}^{(n)})}
{p_{Z^n}(Z^n)},
\notag\\
& R+\eta \geq  \frac{1}{n}\log \ts 
\frac{1}{p_{K^n|M_{\cal A}^{(n)} M^{(n)}}
           (K^n|M_{\cal A}^{(n)},M^{(n)})}
\Biggr\}+6{\rm e}^{-n\eta}, 
\notag
\end{align}
implying that
\begin{align}
& (nR)^{-1}\Upsilon(\varphi_{\cal A}^{(n)},R|
p^n_{\overline{K}\,\overline{Z}})
\notag\\
& \leq p_{M_{\cal A}^{(n)}Z^nK^n}\Biggl\{
\eta  \geq  \frac{1}{n} \log \ts \frac{ q_{{Z}^n}(Z^n)}{p_{Z^n}(Z^n)},
\label{eqn:asppa}\\
& \eta  \geq  \frac{1}{n}\log \ts
 \frac{\hat{q}_{M_{\cal A}^{(n)}Z^nK^n}(M_{\cal A}^{(n)},Z^n,K^n)}
{p_{M_{\cal A}^{(n)}Z^nK^n}(M_{\cal A}^{(n)},Z^n,K^n)},
\label{eqn:asppb}\\
& R_{\cal A}+\eta \geq  \frac{1}{n}\log \ts 
\frac{p_{Z^n|M_{\cal A}^{(n)}}(Z^n|M_{\cal A}^{(n)})}
{p_{Z^n}(Z^n)},
\notag\\
& R+\tau_n +3\eta\geq  \frac{1}{n}\log \ts
\frac{1}{p_{K^n|M_{\cal A}^{(n)}}
           (K^n|M_{\cal A}^{(n)})}
\Biggr\}+6{\rm e}^{-n\eta}. 
\label{eqn:azsad}
\end{align}
In (\ref{eqn:asppa}), we can choose any 
distribution $q_{Z^n}$ on ${\cal Z}^n$. 
In (\ref{eqn:asppb}), we can choose any probability 
distribution $\hat{q}_{M_{\cal A}^{(n)}}$ ${}_{Z^nK^n}$ on 
${\cal M}_{\cal A}^{(n)}$$\times{\cal Z}^n$$\times{\cal X}^n$. 
\end{lemma}

Proof of this lemma is given in Appendix \ref{apd:Apda}.
\begin{remark}
Lemma \ref{lem:Ohzzz}  can be regarded as {\it a new information spectrum meta converse lemma}. In this lemma the bounds (\ref{eqn:NewboundOne}), (\ref{eqn:NewboundTwo}) in $p_{M_{\cal A}^{(n)}M^{(n)}Z^nK^n}\{\cdot \}$ 
are new. An idea of adding those two quantities in $p_{M_{\cal A}^{(n)}M^{(n)}Z^nK^n}\{\cdot \}$ is quite essential to derive    
the bound (\ref{eqn:azsad}).  
\end{remark}

The upper bound (\ref{eqn:azsad}) of 
$(nR)^{-1}\Upsilon(R,$ $\varphi_{\cal A}^{(n)}$ $|p^n_{\overline{K}\,\overline{Z}})$ 
in Lemma \ref{lem:Ohzzz} is the similar to that of 
the correct probability of 
decoding for one helper source coding problem in Lemma 1 in 
Oohama \cite{oohama:19}. In a manner quite similar 
to the derivation of the exponential upper 
bound of the correct probability of decoding for one helper 
source coding problem \cite{oohama:19}, we can derive 
an exponential upper bound of 
$\Upsilon(\varphi_{\cal A}^{(n)},R|
p^n_{\overline{K}\,\overline{Z}})$.
This result is shown in the following proposition.  
\begin{proposition}\label{pro:ThetaExpUpper} 
For any 
$p_{\overline{K}\,\overline{Z}}$ 
$ \in {\cal P}_n({\cal X} \times {\cal Z})$ 
and for any $\varphi_{\cal A}^{(n)} 
\in {\cal F}_{\cal A}^{(n)}(R_{\cal A})$, 
we have 
\begin{align}
& {\Upsilon(\varphi_{\cal A}^{(n)},R|
p^n_{\overline{K}\,\overline{Z}})}
\leq 
7(nR) {\ExP}^{-n G(R_{\cal A},R+\tau_n|
p_{\overline{K}\,\overline{Z}})}
\notag\\
&\leq 
7(nR){\ExP}^{(1/5)n\tau_n} {\ExP}^{-n G(R_{\cal A},R|
p_{\overline{K}\,\overline{Z}})},
\end{align}
where the last inequality follows from Property 
\ref{pr:PrExpG} part a).  
\end{proposition}

{\it Proof of Proposition \ref{pro:KeyPro}:}
From Lemma \ref{lem:Ohzzza} and Proposition \ref{pro:ThetaExpUpper},
we immediately obtain Proposition \ref{pro:KeyPro}.
\hfill\IEEEQED

\newcommand{\Apda}{
\subsection{
Proof of Lemma \ref{lem:Ohzzz}
}\label{apd:Apda}

To prove Lemma \ref{lem:Ohzzz}, we prepare a lemma.
For simplicity of notation, set 
$|{\cal M}_{\cal A}^{(n)}|=M_{\cal A}$.
Define
\begin{align*}
{\cal E}_{1,n} \defeq & \{(a,b,z^{n},k^{n}):
{p_{Z^n}(z^n)}\geq  {\ExP}^{-n\eta}  {q_{Z^n}(z^n)}\},
\\
{\cal E}_{2,n} \defeq &\{(a,b,z^{n},k^{n}):p_{M_{\cal A}^{(n)}Z^nK^n}(a,z^n,k^n) 
\\
&\quad \geq {\ExP}^{-n\eta} 
\hat{q}_{M_{\cal A}^{(n)}Z^nK^n}(a,z^n,k^n)\},
\\
{\cal E}_{3,n} \defeq &\{(a,b,z^{n},k^{n}):
p_{Z^n|K^n M_{\cal A}^{(n)} M^{(n)}}(z^n|k^n,a,b) 
\\
&\quad \geq {\ExP}^{-n\eta} 
p_{Z^n|K^n M_{\cal A}^{(n)} }
(z^n|k^n,a) \},
\end{align*}
Furthermore, define
\begin{align*}
{\cal E}_{4,n}\defeq & \{(a,b,z^n,k^n): 
   a=\varphi_{\cal A}^{(n)}(z^n),b=P_{k^n,z^n},
  \\ 
& p_{K^nZ^n|M_{\cal A}^{(n)}M^{(n)}}(k^n,z^n|a,b)
 \\
& \leq  |{\cal P}_n ({\cal X}\times {\cal Y})|
  {\rm e}^{n\eta}p_{K^nZ^n|M_{\cal A}^{(n)}}(k^n,z^n|a)\}, 
  \\
 {\cal E}_{5,n}\defeq & \{(a,b,z^n,k^n): 
   a=\varphi_{\cal A}^{(n)}(z^n),\\ 
& p_{Z^n|M_{\cal A}^{(n)}}(z^n|a)\leq 
  M_{\cal A}{\rm e}^{n\eta}p_{Z^n}(z^n)\}, 
\\
{\cal E}_{0,n} \defeq & \{(a,b,z^n,k^n): 
 a=\varphi_{\cal A}^{(n)}(z^n),b=P_{k^n,z^n}\\ 
& p_{K^n|M_{\cal A}^{(n)}M^{(n)}}(k^n|a,b) 
  \geq {\rm e}^{-n(R+\eta)}\}.
\end{align*}
Then we have the following lemma. 
\begin{lemma}\label{lem:zzxa}{ For $i=1,2,\cdots,5$, we have
\begin{align}
& p_{M_{\cal A}^{(n)}M^{(n)}Z^nK^n}({\cal E}_{i,n})
\leq {\rm e}^{-n\eta}.
\label{eqn:Bound}
\end{align}
}
\end{lemma}

{\it Proof:} In \cite{santosoOh:19}, 
the bound (\ref{eqn:Bound})
have already been proved for $i=1,2,5.$ Hence it 
suffices to prove  (\ref{eqn:Bound}) for $i=3,4$. 
We first prove (\ref{eqn:Bound}) for $i=3$.
We have the following chain of inequalities:
\begin{align*}
& p_{M_{\cal A}^{(n)}M^{(n)}Z^nK^n}
({\cal E}_{3,n}^{\rm c})
\\
&=  \sum_{(a,b, z^n,k^n) \in {\cal E}_{3,n}^{\rm c}}
      p_{M_{\cal A}^{(n)}M^{(n)} Z^nK^n}(a,b, z^n,k^n)
\\
&=  \sum_{(a,b, z^n,k^n) \in {\cal E}_{3,n}^{\rm c}}
      p_{M_{\cal A}^{(n)}M^{(n)} K^n}(a,b,k^n)
\\
& \qquad\qquad \times
  p_{Z^n|M_{\cal A}^{(n)}M^{(n)} K^n} (z^n|a,b,k^n)  
\\
&\MLeq{a}\sum_{(a,b, z^n,k^n)\in {\cal E}_{3,n}^{\rm c}}
 p_{M_{\cal A}^{(n)}M^{(n)} K^n}(a,b,k^n)
\\
& \qquad\qquad \times {\rm e}^{-n\eta} 
    p_{Z^n|M_{\cal A}^{(n)} K^n} (z^n|a,k^n) 
\\
&\leq\sum_{(a,b,z^n,k^n)}p_{M_{\cal A}^{(n)}M^{(n)} K^n}(a,b,k^n)
\\
& \qquad\qquad \times {\rm e}^{-n\eta} 
   p_{Z^n|M_{\cal A}^{(n)} K^n} (z^n|a,k^n) 
\\
&
={\rm e}^{-n\eta}.
\end{align*}
Step (a) follows from the definition of ${\cal E}_{3,n}$.
We next prove (\ref{eqn:Bound}) for $i=4$.
We have the following chain of inequalities:   
\begin{align*}
& p_{M_{\cal A}^{(n)}M^{(n)}Z^nK^n}({\cal E}_{4,n}^{\rm c})
\\
&=\sum_{b \in {\cal P}_n({\cal X}\times{\cal Z})}
  \sum_{\scs (a,z^n,k^n): P_{z^n,k^n}=b
     \atop{\scs
        p_{K^nZ^n|M_{\cal A}^{(n)}}(k^n,z^n|a) 
           \atop{\scs 
              \leq {\rm e}^{-n\eta}
              |{\cal P}_n({\cal X}\times{\cal Z})|^{-1}
                \atop{\scs
                    \times 
                     p_{K^nZ^n| M_{\cal A}^{(n)}M^{(n)} }(k^n,z^n|a,b)
                     }
               }
         }
     }
1 
\\
& \qquad \times p_{K^nZ^n|M_{\cal A}^{(n)}}(k^n,z^n|a)p_{M_{\cal A}^{(n)}}(a)
\\
&\leq  \frac{{\rm e}^{-n\eta}}
        {|{\cal P}_n({\cal X}\times{\cal Z})|}
\sum_{b \in {\cal P}_n({\cal X}\times{\cal Z})} 
 \sum_{\scs (a,z^n,k^n): P_{z^n,k^n}=b
 \atop{\scs
      p_{K^nZ^n|M_{\cal A}^{(n)}}(k^n,z^n|a) 
      \atop{\scs 
           \leq {\rm e}^{-n\eta}
          |{\cal P}_n({\cal X}\times{\cal Z})|^{-1}
            \atop{\scs
               \times p_{K^nZ^n| M_{\cal A}^{(n)}M^{(n)} }(k^n,z^n|a,b)
                }
           }
     }
}1
\\
&\qquad\times 
p_{K^nZ^n|M_{\cal A}^{(n)}M^{(n)}}(k^n,z^n|a,b)p_{M_{\cal A}^{(n)}}(a)
\\
&\leq  \frac{{\rm e}^{-n\eta}}
        {|{\cal P}_n({\cal X}\times{\cal Z})|}
\sum_{b \in {\cal P}_n({\cal X}\times{\cal Z})} 
 \sum_{\scs (a,z^n,k^n)}1
\\
&\qquad\times 
p_{K^nZ^n|M_{\cal A}^{(n)}M^{(n)}}(k^n,z^n|a,b)p_{M_{\cal A}^{(n)}}(a)
\\
&=\frac{{\rm e}^{ -n\eta }}{|{\cal P}_n({\cal X}\times{\cal Z})|} 
|{\cal P}_n({\cal X}\times{\cal Z})|={\rm e}^{-n\eta},
\end{align*}
completing the proof.
\hfill\IEEEQED

{\it Proof of Lemma \ref{lem:Ohzzz}:} 
By definition we have
\begin{align}
& p_{M_{\cal A}^{(n)}M^{(n)}Z^nK^n}
\left(\bigcap_{i=0}^5{\cal E}_{i,n}\right)
\notag\\
&=p_{M_{\cal A}^{(n)}M^{(n)}Z^nK^n}\Biggl\{
\eta  \geq \frac{1}{n} \log \ts \frac{ q_{{Z}^n}(Z^n)}{p_{Z^n}(Z^n)},
\notag\\
& 
\eta \geq \frac{1}{n}\log \ts 
\frac{\hat{q}_{M_{\cal A}^{(n)}Z^nK^n}(M_{\cal A}^{(n)},Z^n,K^n)}
{p_{M_{\cal A}^{(n)}Z^nK^n}(M_{\cal A}^{(n)},Z^n,K^n)},
\notag\\
& \eta \geq  \frac{1}{n} \log \ts
\frac{p_{Z^n|K^n M_{\cal A}^{(n)}}
        (Z^n|K^n, M_{\cal A}^{(n)})}
     {p_{Z^n|K^n M_{\cal A}^{(n)} M^{(n)}}
        (Z^n|K^n,M_{\cal A}^{(n)},M^{(n)})},
\notag \\
&\frac{1}{n}\log|{\cal P}_n({\cal X}\times {\cal Z})|+\eta 
\notag\\
&\qquad\quad 
\geq  \frac{1}{n}\log \ts 
\frac{p_{K^n Z^n| M_{\cal A}^{(n)}M^{(n)}}
        (K^n,Z^n| M_{\cal A}^{(n)},M^{(n)})}
     {p_{K^n Z^n| M_{\cal A}^{(n)}}
        (K^n,Z^n| M_{\cal A}^{(n)})},
\notag \\
& \frac{1}{n}\log M_{\cal A}+\eta\geq  \frac{1}{n}\log \ts
\frac{p_{Z^n|M_{\cal A}^{(n)}}(Z^n|M_{\cal A}^{(n)})}
{p_{Z^n}(Z^n)},
\notag\\
& R+\eta \geq  \frac{1}{n}\log \ts 
\frac{1}{p_{K^n|M_{\cal A}^{(n)} M^{(n)}}
           (K^n|M_{\cal A}^{(n)},M^{(n)})}
\Biggr\}.
\label{eqn:ZZA}
\end{align}
Here we note that by Lemma \ref{lem:Lem1} part a), we have
\begin{align}
\frac{1}{n}\log|{\cal P}_n({\cal X}\times {\cal Z})|\leq \tau_n.
\label{eqn:ZZAb}
\end{align}
Then from (\ref{eqn:ZZA}) and (\ref{eqn:ZZAb}) we have   
\begin{align}
& p_{M_{\cal A}^{(n)}M^{(n)}Z^nK^n}
\left(\bigcap_{i=0}^5{\cal E}_{i,n}\right)
\notag\\
&\leq p_{M_{\cal A}^{(n)}M^{(n)}Z^nK^n}\Biggl\{
\eta  \geq \frac{1}{n} \log \ts \frac{ q_{{Z}^n}(Z^n)}{p_{Z^n}(Z^n)},
\notag\\
& 
\eta \geq \frac{1}{n}\log \ts 
\frac{\hat{q}_{M_{\cal A}^{(n)}Z^nK^n}(M_{\cal A}^{(n)},Z^n,K^n)}
{p_{M_{\cal A}^{(n)}Z^nK^n}(M_{\cal A}^{(n)},Z^n,K^n)},
\notag\\
& \eta \geq  \frac{1}{n} \log \ts
\frac{p_{Z^n|K^n M_{\cal A}^{(n)}}
        (Z^n|K^n, M_{\cal A}^{(n)})}
     {p_{Z^n|K^n M_{\cal A}^{(n)} M^{(n)}}
        (Z^n|K^n,M_{\cal A}^{(n)},M^{(n)})},
\notag \\
& \tau_n +\eta  
\geq  \frac{1}{n}\log \ts 
\frac{p_{K^n Z^n| M_{\cal A}^{(n)}M^{(n)}}
        (K^n,Z^n| M_{\cal A}^{(n)},M^{(n)})}
     {p_{K^n Z^n| M_{\cal A}^{(n)}}
        (K^n,Z^n| M_{\cal A}^{(n)})},
\notag \\
& \frac{1}{n}\log M_{\cal A}+\eta\geq  \frac{1}{n}\log \ts
\frac{p_{Z^n|M_{\cal A}^{(n)}}(Z^n|M_{\cal A}^{(n)})}
{p_{Z^n}(Z^n)},
\notag\\
& R+\eta \geq  \frac{1}{n}\log \ts 
\frac{1}{p_{K^n|M_{\cal A}^{(n)} M^{(n)}}
           (K^n|M_{\cal A}^{(n)},M^{(n)})}
\Biggr\}. 
\end{align}
Then for any $\varphi_{\cal A}^{(n)}$ 
satisfying 
$
(1/n)\log {||\varphi_{\cal A}^{(n)}||} \leq R_{\cal A},
$
we have 
\begin{align}
& p_{M_{\cal A}^{(n)}M^{(n)}Z^nK^n}
\left(\bigcap_{i=0}^5{\cal E}_{i,n}\right)
\notag\\
&\leq p_{M_{\cal A}^{(n)}M^{(n)}Z^nK^n}\Biggl\{
\eta  \geq \frac{1}{n} \log \ts \frac{ q_{{Z}^n}(Z^n)}{p_{Z^n}(Z^n)},
\notag\\
& 
\eta \geq \frac{1}{n}\log \ts 
\frac{\hat{q}_{M_{\cal A}^{(n)}Z^nK^n}(M_{\cal A}^{(n)},Z^n,K^n)}
{p_{M_{\cal A}^{(n)}Z^nK^n}(M_{\cal A}^{(n)},Z^n,K^n)},
\notag\\
& \eta \geq  \frac{1}{n} \log \ts
\frac{p_{Z^n|K^n M_{\cal A}^{(n)}}
        (Z^n|K^n, M_{\cal A}^{(n)})}
     {p_{Z^n|K^n M_{\cal A}^{(n)} M^{(n)}}
        (Z^n|K^n,M_{\cal A}^{(n)},M^{(n)})},
\notag \\
& \tau_n +\eta  
\geq  \frac{1}{n}\log \ts 
\frac{p_{K^n Z^n| M_{\cal A}^{(n)}M^{(n)}}
        (K^n,Z^n| M_{\cal A}^{(n)},M^{(n)})}
     {p_{K^n Z^n| M_{\cal A}^{(n)}}
        (K^n,Z^n| M_{\cal A}^{(n)})},
\notag \\
& R_{\cal A}+\eta\geq  \frac{1}{n}\log \ts
\frac{p_{Z^n|M_{\cal A}^{(n)}}(Z^n|M_{\cal A}^{(n)})}
{p_{Z^n}(Z^n)},
\notag\\
& R+\eta \geq  \frac{1}{n}\log \ts 
\frac{1}{p_{K^n|M_{\cal A}^{(n)} M^{(n)}}
           (K^n|M_{\cal A}^{(n)},M^{(n)})}
\Biggr\}.
\end{align}
Hence, it suffices to show 
\begin{align*}
&  \wp^{(n)}_{\eta}(\varphi_{\cal A}^{(n)},R|p_{KZ}^n) 
\leq p_{M_{\cal A}^{(n)}M^{(n)}Z^nK^n}
\left(\bigcap_{i=0}^5{\cal E}_{i,n}\right)
+5{\rm e}^{-n\eta}.
\end{align*}
to prove Lemma \ref{lem:Ohzzz}. We have the following 
chain of inequalities:
\begin{align*}
& \wp^{(n)}_{\eta}(\varphi_{\cal A}^{(n)},R|p_{KZ}^n) 
  = p_{M_{\cal A}^{(n)}M^{(n)}Z^nK^n}\left({\cal E}_{0,n}\right)
\\
&= p_{M_{\cal A}^{(n)}M^{(n)}Z^nK^n}
   \left(\bigcap_{i=0}^5{\cal E}_{i,n}\right)
\\
&\quad  +p_{M_{\cal A}^{(n)}M^{(n)}Z^nK^n}
  \left(
 \left[\bigcap_{i=1}^5{\cal E}_{i,n}\right]^{\rm c} 
 \cap {\cal E}_{0,n}
 \right)
\\
&\leq p_{M_{\cal A}^{(n)}M^{(n)}Z^nK^n}
   \left(\bigcap_{i=0}^5{\cal E}_{i,n}\right)
\\
&\quad +\sum_{i=1}^5
    p_{M_{\cal A}^{(n)}M^{(n)}Z^nK^n}
   \left({\cal E}_{i,n}^{\rm c}\right)
\\
&\MLeq{a}
p_{M_{\cal A}^{(n)}M^{(n)}Z^nK^n}
   \left(\bigcap_{i=0}^5{\cal E}_{i,n}\right)
+5{\rm e}^{-n\eta}. 
\end{align*}
Step (a) follows from Lemma \ref{lem:zzxa}.
\hfill\IEEEQED
}

\Apda

\bibliographystyle{IEEEtran}
\bibliography{Isit2024.bib}

\end{document}